\documentstyle[preprint,prd,aps,psfig]{revtex}

\newcommand{\rsub}[1]{\mbox{\tiny #1}}

\begin{document}

\draft
\preprint{\vbox{\hbox{HUPD-9720}}}

\title{
 Lattice study of $B\rightarrow \pi$ semileptonic decay \\
 using nonrelativistic lattice QCD  }

\author{Shoji Hashimoto
  \footnote{permanent address:
    Computing Research Center,
    High Energy Accelerator Research Organization (KEK),
    Tsukuba 305-0801, Japan}}
\address{Theoretical Physics Department,
         Fermi National Accelerator Laboratory,\\
         P.O. Box 500, Batavia, IL60510}
\author{Ken-Ichi Ishikawa, Hideo Matsufuru, Tetsuya Onogi,
        and Norikazu Yamada}
\address{Department of Physics, Hiroshima University,
         Higashi-Hiroshima 739-8526, Japan }

\date{November 1997, \ revised March 1998}

\maketitle

\begin{abstract}
  \setlength{\baselineskip}{3ex}
  We present an exploratory lattice study of $B\rightarrow\pi$
  semileptonic decay form factors using the nonrelativistic 
  lattice QCD for heavy quark with Wilson light quark
  on a $16^3 \times 32$ quenched lattice at $\beta=5.8$.
  The matrix elements are calculated at eight values of heavy
  quark mass in a range of $1.5-8$ GeV and with three values
  of light quark mass.
  The $1/m_B$ corrections to the matrix elements are found to
  be fairly small except for the spatial component 
  proportional to the $B$ meson momentum.
  We find that the $q^2$ dependence of the form factor $f^+(q^2)$
  near $q_{\rsub{max}}^2$ becomes much stronger for larger heavy quark mass, 
  which may suggest the increase of a pole contribution.  
  We perform a model independent fit of the form factors
  and study whether the $q^2$ dependence is consistent 
  with the pole contribution.
  Although the soft pion theorem predicts
  $f^0(q_{\rsub{max}}^2)=f_B/f_{\pi}$ in the chiral limit, we 
  observe a significant violation of this relation.
\end{abstract}


\newpage
\setlength{\baselineskip}{3ex}

\section{INTRODUCTION}
\label{sec:INTRODUCTION}

The exclusive semileptonic decay 
$B\rightarrow\pi(\rho)l \nu$
would become an important process to determine the
Cabibbo-Kobayashi-Maskawa (CKM) matrix element $|V_{ub}|$,
when high statistics experimental data becomes available by
future $B$ Factories, since
the lattice QCD simulation enables us to compute the
relevant form factors from the first principle. 
There was, however, a difficulty in treating a heavy quark 
with mass $m_Q$ on the lattice, 
because possible systematic errors of order
$am_Q$ could become unacceptably large for a typical lattice
spacing $a$ accessible in present simulations.
Thus the previous lattice calculations of the $B$ meson 
semileptonic decay form factors \cite{APE95,UKQCD95,Wuppertal96}
involved an extrapolation in the heavy quark mass from
the charm quark mass regime to the $b$-quark mass assuming
a heavy quark mass scaling law, which could
introduce a potential systematic error.
The nonrelativistic lattice QCD (NRQCD) \cite{TL91} is 
designed to remove such a large uncertainty based on a 
systematic $1/m_Q$ expansion, and one can simulate the 
$b$-quark directly at its mass value.
In this paper we describe the lattice calculation of 
the $B\rightarrow\pi l\nu$ form factors using the NRQCD for 
heavy quark.
We investigate the heavy quark mass dependence of the form
factors, by taking the mass of the heavy quark to cover
a range of 1.5--8 GeV.

The hadronic matrix elements of the $B \rightarrow \pi$ 
semileptonic decay are expressed in terms of two form 
factors $f^+$ and $f^0$ as
\begin{equation}
\langle \pi(\vec{k}) | V_{\mu} | B(\vec{p}) \rangle =
 \left( p + k - q \frac{ m_B^{\,2} - m_{\pi}^{\,2} }{q^2} 
                   \right)_{\mu}  f^{+}(q^2) 
     + q_{\mu}\frac{ m_B^{\,2} 
                      - m_{\pi}^{\,2} }{q^2} f^{0}(q^2),
\end{equation}
where $q_{\mu}=p_{\mu}-k_{\mu}$.
The covariant normalization of the meson fields is employed
in this paper:
\begin{equation}
\langle M(\vec{p}') | M(\vec{p}) \rangle =
  2 p_0 (2\pi)^3 \delta^3( \vec{p} - \vec{p}' ).
\end{equation}
From the condition that the matrix element is not singular
at $q^2=0$, the form factors satisfy $f^{+}(0)=f^{0}(0)$,
and the kinematical end point
$q^2_{\rsub{max}} =(m_{B}-m_{\pi})^2$
corresponds to the zero recoil limit, where the lattice 
simulation works most efficiently.

This paper is organized as follows.
In the next section, the NRQCD formulation and numerical
method to calculate the matrix elements are summarized.
We describe the details of our numerical simulation in
Section \ref{sec:SIMULATION_DETAILS}, where we point out the
subtleties in extracting the form factors: 
the definition of the heavy meson energy, the choice
of the two independent matrix elements to determine
$f^+$ and $f^0$, and the procedure of chiral extrapolation. 
We explain what we think is the best procedure and study the
uncertainty by comparing the result with those from other
procedures. 
Physical implications of numerical results are discussed in
Section \ref{sec:PHYSICAL_IMPLICATIONS}. 
We study $1/m_B$ dependence of the matrix elements
and $q^2$ dependence of the form factors.
The prediction from the soft pion theorem is compared with
our data.
In Section \ref{sec:SYSTEMATIC_ERRORS}, we discuss the
systematic uncertainties contained in this work.
Section \ref{sec:CONCLUSION} is devoted to our conclusion.

\section{LATTICE NRQCD}
\label{sec:LATTICE_NRQCD}

\subsection{Lattice NRQCD action}

The lattice NRQCD has been extensively used for 
the investigations of the heavy-heavy systems \cite{Shigemitsu97a} 
and hadrons containing a single heavy quark \cite{Onogi97}.
It is designed to remove the large mass scale $m_Q$ from
the theory using the $1/m_Q$ expansion and to reproduce
the same results as of the relativistic QCD up to a given
order of $1/m_Q$. 
In this work, we employ the lattice NRQCD action including
the $O(1/m_Q)$ terms\footnote{This action differs from that
  we used in our previous study of $f_B$ \cite{Hiroshima97},
  which is organized to remove $O(a\Lambda_{QCD}/m_Q)$ error,
  at the cost of simulation speed.}
\begin{equation} 
  S_{\rsub{NRQCD}} 
  = \sum_x Q^{\dag}(x)\left[
      \left(1- \frac{1}{2n}H_0 \right)^{-n}
      U_{4} \left(1- \frac{1}{2n}H_0 \right)^{-n}
      Q(x+\hat{4}) - (1-\delta H )  Q(x)  \right],
  \label{eq:NRQCDaction}
\end{equation}
where
\begin{eqnarray}
  H_0       &=&  -\frac{1}{2m_Q} \Delta^{(2)}, \\
  \delta H  &=&  -\frac{1}{2m_Q} \vec{\sigma}\cdot \vec{B},
\end{eqnarray}
and $Q(x)$ is the effective two component spinor field,
which describes the heavy quark.
$\Delta^{(2)}$ denotes the three dimensional Laplacian,
and $\vec{B}$ is the standard clover-leaf type 
chromomagnetic field.

This action generates the following evolution equation
\begin{eqnarray}
  G_{Q}(\vec{x},t=1) &=& \left( 1-\frac{1}{2n}H_0 \right)^n
      U_4^{\dag} \left( 1-\frac{1}{2n}H_0 \right)^n \, 
      G_{Q}(\vec{x},t=0),
  \label{eq:evolve1} \\
  G_{Q}(\vec{x},t+1) &=& \left( 1-\frac{1}{2n}H_0 \right)^n
      U_4^{\dag} \left( 1-\frac{1}{2n}H_0 \right)^n
      ( 1 - \delta H ) \, G_{Q}(\vec{x},t),
  \label{eq:evolve2}
\end{eqnarray}
for which we apply tadpole improvement procedure
$U_{\mu }(x) \rightarrow U_{\mu }(x)/u_0$ with
$u_0=\langle \mbox{Tr}U_{\rsub{plaq}}/3\rangle^{1/4}$
\cite{LM93}.
To avoid the singular behavior of high frequency modes
in the evolution equation,
the stabilizing parameter $n$ is chosen to satisfy
the condition $|1-H_0/2n|<1$, which leads to $n>3/2m_Q$.
From a viewpoint of the perturbation theory,
further constraint, $H_0/2n<1$, is necessary to avoid
singularities in some of vertices derived
from the action (\ref{eq:NRQCDaction}).
This point is again discussed in the last part of this
section in connection with our choice of $n$ 
in the simulation and the perturbative calculation.

The four component spinor field $\psi(x)$ of the 
relativistic QCD is expressed in terms of two component
spinor field $Q(x)$ as
\begin{equation}
  \psi(x) =
    \left( 1 - \frac{\vec{\gamma}\cdot\vec{\Delta}}{2m_Q}
                                                   \right)
    \left( \begin{array}{c} Q(x) \\ 0 \end{array} \right),
  \label{eq:psi_field}
\end{equation}
where $\vec{\Delta}$ is the symmetric lattice covariant
derivative. 
$O(1/m_Q)$ correction appears in the lower component of 
$\psi(x)$, which affects the heavy-light current.

\subsection{Correlation functions}

We employ the standard simulation technique to calculate the
hadronic matrix elements of the semileptonic decay.
We calculate the three-point correlation functions
\begin{equation}
  C_{\mu}^{(3)}(\vec{p},\vec{k};t_{\!f},t_{\!s},t_{\!i})
  = \sum_{\vec{x}_{\!f}} \sum_{\vec{x}_{\!s}}
   e^{-i\vec{p}\cdot\vec{x}_{\!f}}
   e^{-i(\vec{k}-\vec{p}) \cdot\vec{x_{\!s}}}
     \, \langle 0 | O_B( \vec{x}_{\!f},t_{\!f} )
        V^{ \dag}_{\mu}(\vec{x}_{\!s},t_{\!s} )
        O_{\pi}^{\dag}(\vec{0},t_{\!i}) | 0 \rangle,
  \label{eq:threept}
\end{equation}
where $O_B$ and $O_{\pi}$ are the interpolating operators
for $B$ and $\pi$ mesons, respectively,
and $V_{\mu}=\bar{q}\gamma_{\mu}\psi$ is the heavy-light 
vector current.
In this work, we use the Wilson quark to describe light
quark $q(x)$.
Here we denote the heavy-light and the light-light 
pseudoscalar mesons as $B$ and $\pi$, respectively, 
regardless of their mass parameters $\kappa$ and $m_Q$ 
for simplicity.
For $t_{\!f} \gg t_{\!s} \gg t_{\!i}$ the correlation
function Eq. (\ref{eq:threept}) becomes
\begin{equation}
  C_{\mu}^{(3)}(\vec{p},\vec{k};t_{\!f},t_{\!s},t_{\!i})
  \rightarrow
  \frac{Z_B(\vec{p})}{2 E_B(\vec{p})} 
  \frac{Z_{\pi}(\vec{k})}{2 E_{\pi}(\vec{k})}
  e^{-E_{\bar{q}Q}(\vec{p})(t_{\!f} - t_{\!s})}
  e^{-E_{\pi}     (\vec{k})(t_{\!s} - t_{\!i})}
   \langle B(\vec{p}) | V^{ \dag}_{\mu} | \pi(\vec{k}) 
   \rangle_{\rsub{latt}},
  \label{eq:3point_func}
\end{equation}
where $E_B(\vec{p})$ and $E_{\pi}(\vec{k})$ denote the energy
of $B$ meson and pion, respectively.
The exponent $E_{\bar{q}Q}(\vec{p})$ is not the total energy
but the binding energy of the $B$ meson, because the heavy
quark mass $m_Q$ is subtracted in the NRQCD.
We use the local interpolating operators for both of $B$
and $\pi$, and
\begin{equation}
  Z_B(\vec{p})     = \langle 0|O_B(0)|B(\vec{p}) \rangle,
  \mbox{\hspace*{1cm}}
  Z_{\pi}(\vec{k}) = \langle 0|O_{\pi}(0)|\pi(\vec{k})\rangle
\end{equation}
are their matrix elements.

In calculating Eq.~(\ref{eq:threept})
we vary $t_f$ with fixed $t_i$ and $t_s$ in order to find out the region
where the correlation functions are dominated by the ground
state.
The fixed $t_s$ is chosen so that the pion two-point correlation 
function is dominated by the ground state, as is shown 
in Section \ref{sec:SIMULATION_DETAILS}.
To obtain $E_{\bar{q}Q}(\vec{p})$, $E_{\pi}(\vec{k})$,
$Z_B(\vec{p})$ and $Z_{\pi}(\vec{k})$,
we also calculate the two-point correlation functions
with a finite momentum
\begin{eqnarray}
  C^{(2)}_{B}(\vec{p};t_f,t_i)
  &=&  \sum_{\vec{x}_f} e^{-i\vec{p}\cdot \vec{x}_f}
       \langle O_{B}(x_f) O_{B}^{\dag}(x_i)  \rangle
  \rightarrow  \frac{Z_B(\vec{p})^2}{2E_{B}(\vec{p})}
       \exp(-E_{\bar{q}Q}(\vec{p}) ( t_f - t_i ) ),
  \label{eq:B_2point} \\
  C^{(2)}_{\pi}(\vec{k};t_f,t_i)
  &=&  \sum_{\vec{x}_f} e^{-i\vec{k}\cdot \vec{x}_f}
       \langle O_{\pi}(x_f) O_{\pi}^{\dag}(x_i)  \rangle
  \rightarrow  \frac{Z_{\pi}(\vec{k})^2}{2E_{\pi}(\vec{k})}
              \exp(-E_{\pi}(\vec{k}) ( t_f - t_i ) ).
\label{eq:pi_2point}
\end{eqnarray}
Combining Eqs.(\ref{eq:3point_func}), (\ref{eq:B_2point}),
and (\ref{eq:pi_2point}),
one can easily see that the matrix element is expressed as
\begin{equation}
  \langle B(\vec{p})| V_{\mu}^{\dag} | \pi(\vec{k})
                                  \rangle_{\rsub{latt}}
  = \sqrt{2 E_B(\vec{p})} \sqrt{2 E_{\pi}(\vec{k})}
   \frac{ e^{E_{\bar{q}Q}(t_f-t_s)} }{\tilde{Z}_B(\vec{p})}
   \frac{ \tilde{Z}_{\pi}(\vec{k}) }
                    { C_{\pi}^{(2)}(\vec{k};t_s,t_i) }
   C^{(3)}_{\mu}(\vec{p},\vec{k} ; t_f,t_s,t_i)
  \label{eq:V_latt}
\end{equation}
for $t_{\!f} \gg t_{\!s} \gg t_{\!i}$, 
where $\tilde{Z}=Z/\sqrt{2E}$.
As expressed in Eq. (\ref{eq:V_latt}), we use the two-point
correlation function itself to cancel the exponentially
decaying factor of pion, while use the values of 
$E_{\bar{q}Q}$ obtained by fits to cancel the $B$ meson's.
One reason of this asymmetric procedure is that 
the pion two-point function is constructed from the light
quark propagator with a point source at $t_i=4$, which is
what we used to calculate the three-point function
(\ref{eq:threept}), and then we expect the statistical
fluctuation mostly cancels between (\ref{eq:threept}) and
(\ref{eq:pi_2point}), while for the $B$ meson exponential
function, such a cancellation is not expected.
In addition, as we mention in the next section, 
the two-point correlation function of $B$ meson with the
point source (\ref{eq:B_2point}) requires larger time
separation to reach the plateau than the three-point
function (\ref{eq:threept}), for which the heavy quark
source is effectively `smeared' at $t_s$.

\subsection{Perturbative corrections}
\label{subsec:Pcorr}

To relate the matrix element in the lattice theory to that
in the continuum QCD, operator matching is required.
We have calculated the perturbative renormalization factor
$Z_{V_{\mu}}$ for the vector current at one-loop level
using the lattice perturbation theory \cite{Ishikawa97}
\begin{equation}
V_{\mu}^{\mbox{\tiny cont}} 
  = Z_{V_{\mu}} \; V_{\mu}^{\mbox{\tiny latt}}
  = Z_{V_{\mu}} \; \bar{q}\gamma_{\mu} \psi,
\label{eq:cont_latt}
\end{equation}
where $q$ is the Wilson light quark and $\psi$ is defined
in Eq. (\ref{eq:psi_field}).
$Z_V$ is the ratio of the on-shell S-matrix elements in the
continuum theory with $\overline{\mbox{MS}}$ scheme and that in the
lattice theory.
In our definition, $Z_V$ contains the leading logarithmic term,
$\alpha \log(m_Qa)$, which comes from the continuum renormalization
factor.

In calculating $Z_V$ we use the massless Wilson quark and the 
external momenta are taken to be zero. 
We did not take into account the one-loop operator mixing
with higher derivative operators,
since there are already $O(a)$ errors at tree-level 
from the Wilson quark action.
The one-loop coefficient
is modified with the tadpole improvement \cite{LM93}.
For the mean link variable we use 
$u_0=\langle Tr U_{\rsub{plaq}}/3\rangle^{1/4}$
except for the light quark wave function renormalization,
for which we use another possible definition, $u_0=1/8\kappa_c$
\cite{Lepage92}. 
Their one-loop perturbative expressions are used
to determine the perturbative coefficients of $Z_{V_{\mu}}$.

The results for the one-loop coefficient $C_{V_{\mu}}$ in
\begin{equation}
  Z_{V_{\mu}} = 1 + g^2 C_{V_{\mu}}
\end{equation}
are presented in Table \ref{tab:P_corr} for several values
of $(m_Q,n)$.
These values contain the leading logarithmic contribution,
$\log(m_Qa)/4\pi^2$.
The values of $Z_{V_{\mu}}$ with two choices of the lattice
coupling constant $g_V^2(\pi/a)=2.19$ and
$g_V^2(1/a)=3.80$ are plotted 
as a function of $1/m_Q$ in Fig.~\ref{fig:Z_V}.
We observe that the spatial component of the vector 
current receives larger perturbative corrections than
the temporal one.
On the other hand, the $1/m_Q$ dependence is rather stronger
for $Z_{V_4}$ than for $Z_{V_i}$.

When we discuss the $1/m_B$ dependence of the renormalized
matrix elements in Section \ref{sec:PHYSICAL_IMPLICATIONS},
we multiply the leading logarithmic factor
\begin{equation}
 \Theta(m_B/m_B^{\rsub{(phys)}}) 
  = \left( \frac{\alpha_V(m_B)}
         {\alpha_V(m_B^{\rsub{(phys)}})} \right)^{2/11}
  \label{eq:Theta}
\end{equation}
to cancel the logarithmic divergence in the infinite heavy
quark mass limit due to the anomalous dimension of the
heavy-light current. 

The perturbative correction for the heavy quark self-energy 
is also calculated, and the $B$ meson mass is given through
the binding energy of the heavy-light meson
$E_{q\bar{Q}}(\vec{p}=0)$ as
\begin{equation}
 m_{B} = Z_m m_Q -E_0 + E_{\bar{q}Q}(\vec{p}=0),
\label{eq:B_mass}
\end{equation}
where the energy shift $E_0$ and the mass renormalization
$Z_m$ are obtained perturbatively
\begin{eqnarray}
  Z_m & = & 1 + g^2 B, \\
  E_0 & = & g^2 A.
\end{eqnarray}
The tadpole improved coefficients $A$ and $B$ are
also given in Table \ref{tab:P_corr}.

For a historical reason, the stabilizing parameter we have
used does not always satisfy the condition $n>3/m_Q$,
which is necessary to avoid divergent tree level vertices,
while the simulation itself is stable with the condition
$n>3/2m_Q$.
We, therefore, quote the results at tree level in the later
sections as our main results.
We estimate the size of the renormalization effect
with the one-loop coefficients obtained with the combinations
of $m_Q$ and $n$, for which $n$'s are larger than those we have
used in the simulation and the perturbation theory exists.
Although this estimation is certainly incorrect,
it gives some idea for the one-loop effect,
especially because the $n$-dependence of the simulation results
is observed to be very small (Section~\ref{subsec:3pt_func}).

\section{SIMULATION DETAILS}
\label{sec:SIMULATION_DETAILS}

In this section, we describe the numerical simulation in
detail apart from discussions on physical implications of
the results, which will be discussed in the next section.
After summarizing the simulation parameters,
two-point correlation functions of $\pi$ and $B$ mesons
with finite momenta are discussed.
We describe how to extract the matrix elements and the
form factors from the three-point correlation functions.
Finally, the chiral extrapolation of the matrix element is
discussed.

\subsection{Simulation parameters}

The numerical simulations are performed on a 
$16^3 \times 32$ lattice with 120 quenched gauge 
configurations generated with the standard plaquette
gauge action at $\beta$=5.8.
Each configuration is separated by 2000 pseudo-heat-bath
sweeps after 20000 sweeps for thermalization and fixed to
the Coulomb gauge.
The Wilson quark action is used for the light quark at
three $\kappa$ values 0.1570, 0.1585 and 0.1600, which 
roughly lie in the range $[m_s,2m_s]$, and the critical
hopping parameter is $\kappa_c$=0.16346(7).
The boundary condition for the light quark is periodic
and Dirichlet for spatial and temporal directions,
respectively.
The light quark field is normalized with the tadpole
improved form $\sqrt{1-3 \kappa/4 \kappa_c}$
according to \cite{Lepage92}.
The tadpole improvement is also applied for both
the NRQCD action and the current operator 
with the replacement of $U_{\mu} \rightarrow U_{\mu}/u_0$ 
using the average value of a single plaquette 
$u_0=\langle \mbox{Tr}U_{\rsub{plaq}}/3\rangle^{1/4}=0.867994(13)$.

The lattice scale is determined from the $\rho$ meson mass
as $a^{-1}$=1.71(6)~GeV, although we expect a large $O(a)$
error for $m_{\rho}$ with the unimproved Wilson fermion.
The results for the $\pi$ and the $\rho$ meson masses and the
pion decay constant are summarized 
in Table \ref{tab:Spc_pi}.

The heavy quark mass $m_Q$ and the stabilizing parameter
$n$ used in our simulation are
\begin{equation}
  \left( \begin{array}{c}  m_{Q} \\  n  \end{array} \right)
= \left( \begin{array}{c}   5.0  \\  1  \end{array} \right),
  \left( \begin{array}{c}   2.6  \\  1  \end{array} \right),
  \left( \begin{array}{c}   2.1  \\  1  \end{array} \right),
  \left( \begin{array}{c}   2.1  \\  2  \end{array} \right),
  \left( \begin{array}{c}   1.5  \\  2  \end{array} \right),
  \left( \begin{array}{c}   1.2  \\  2  \end{array} \right),
  \left( \begin{array}{c}   1.2  \\  3  \end{array} \right),
  \left( \begin{array}{c}   0.9  \\  2  \end{array} \right),
\end{equation}
where $m_Q=2.6$ and $0.9$ roughly correspond to 
$b$- and $c$-quark masses, respectively.

For $m_Q$=2.1 and 1.2 we performed two sets of simulations
with different values of $n$, though the statistics is
lower (=60) for $(m_Q,n)=(2.1,2)$ and $(1.2,3)$.
Since the different choice of $n$ introduces
the different higher order terms in $a$ in the evolution 
equation,
the choice of $n$ should not affect the physical results
for sufficiently small $a$.
The small dependence of the numerical results on $n$
is also crucial for our estimation of the perturbative
corrections.

The spatial momentum of the $B$ meson ($\vec{p}$) and 
the pion ($\vec{k}$) is taken up to
$|\vec{p}|$, $|\vec{k}| \; \leq \sqrt{3} \cdot 2\pi/16$,
which corresponds to the maximum momentum of $\sim$ 1.2~GeV
in the physical unit.
We measure the three-point correlation function at 20
different momentum configurations $(\vec{p},\vec{k})$ 
as listed in Table \ref{tab:momset}.
The momentum configurations which are equivalent under the
lattice rotational symmetry are averaged, and the number of
such equivalent sets are also shown in 
Table \ref{tab:momset}.

The light quark propagator is solved with a local source at
$t_i$=4, which is commonly used for the two-point and
three-point functions.
The heavy-light vector current is placed at $t_s=14$,
which is chosen so that the pion correlation function is
dominated by the ground state signal.
The position of the $B$ meson interpolating operator is
varied in a range $t_f=23-28$, where we observe a good
plateau as shown later.

\subsection{Light-light meson}

In order to obtain the form factors reliably, it is crucial
to extract the ground state of the $B$ meson and the pion
involving finite momentum properly.
In Fig.~\ref{fig:EP_pi} we show the effective mass plot
of pions with finite momentum at $\kappa=0.1570$ and $0.1600$.
The spatial momentum $\vec{k}=(k_x,k_y,k_z)$ is 
understood with the unit of $2\pi/16$.
This notation will be used throughout this paper.
Although higher momentum states are rather noisy, we
can observe a plateau beyond $t=14$. 
We fit the data with the single exponential function to
obtain the energy $E_{\pi}(\vec{k})$ shown by the horizontal
solid lines in Fig. \ref{fig:EP_pi}.

Figure \ref{fig:DR_pi} shows the energy momentum dispersion
relation of pion, where the solid lines represent the
relation in the continuum
$E_{\pi}(\vec{k})^2=m_{\pi}^2 + \vec{k}^2$.
We observe a small discrepancy between the above relation
and the data, which indicates the discretization error.
However the disagreement is about 1--1.5 standard deviation
and only a few percent.

\subsection{Heavy-light meson}
\label{subsec:Hlmeson}

To compute the $B$ meson two-point correlation
functions, we employ the smeared source for heavy quark as
well as the local source, with the local sink for both
cases.
The smearing function for the heavy quark is obtained with
a prior measurement of the wave function with the local
source.
In Fig. \ref{fig:EP_B} we plot the effective mass for both
the local-local and the smeared-local correlation functions at
$m_Q=2.6$ and $\kappa=0.1570$, $0.1600$.
The plateau is reached beyond $t=16$ for the local-local,
while the smeared-local exhibits clear plateau from even
earlier time slices.
We obtain the binding energy with a fit range $[16,24]$
for both types of the correlation functions and for all
momenta, and the results are consistent in all cases.
The binding energy averaged over the results fitted
from the local and the smeared sources
are listed in Table \ref{tab:Spc_B} together with the
values in the chiral limit.
In Table \ref{tab:Spc_B}, we also listed the binding energy
for the vector meson $B^*$ measured with the local-local
correlation function, which are used in later discussions
on the $B^*$ pole contribution to the form factors.
It is also worth to note that the values of $E_{\bar{q}Q}$
obtained with different stabilizing parameter $n$ is 
consistent with each other within their statistical errors.

The dispersion relation for the $B$ meson takes the
following nonrelativistic form 
\begin{equation}
  E_{\bar{q}Q}(\vec{p}) = E_{\bar{q}Q}(0)
       + \frac{1}{2 m_{\rsub{kin}}} \vec{p}^2
       + O(1/m_B^{\,3}),
\label{eq:DR_B}
\end{equation}
where the kinetic mass $m_{\rsub{kin}}$ should agree with the rest
mass $m_B$ (\ref{eq:B_mass}) in the continuum limit.
Since we use the NRQCD action correct up to $O(1/m_Q)$,
including higher order terms in $1/m_B$ in
Eq.~(\ref{eq:DR_B}) does not make sense.
In Fig. \ref{fig:DR_B}, $E_{\bar{q}Q}(\vec{p})$ is shown as a
function of $\vec{p}^2$ at $m_Q=2.6$.
The solid lines represent the relation (\ref{eq:DR_B}) with
$m_{\rsub{kin}}=m_B$ determined through the tree level relation
$m_B=m_Q+E_{\bar{q}Q}(0)$, which reproduce the data quite well.
With the one-loop correction (\ref{eq:B_mass}) the
agreement becomes even better as presented with the
dashed lines in the figure.

\subsection{Three-point function and matrix elements}
\label{subsec:3pt_func}

Figure \ref{fig:EP_C3} is the effective mass plot of the
three-point function at $m_Q=2.6$ and $\kappa=0.1570$, $0.1600$.
The horizontal axis represents the time slice on which the
$B$ meson interpolating operator is put, and
the vertical axis corresponds to the binding energy of the
$B$ meson.
The horizontal solid lines represent the binding energy
$E_{\bar{q}Q}(\vec{p})$ determined from the two-point
correlation functions.
The figures display that the three-point correlation
functions are dominated by the ground states beyond
$t=23$, and there they give the consistent values for
$E_{\bar{q}Q}(\vec{p})$ with ones extracted
from the two-point functions.
Therefore, in this region we can use
Eq.~(\ref{eq:V_latt}) together with the results of the
two-point correlation functions
to extract the matrix elements.

It is useful to define the quantity $\hat{V}_{\mu}$ as
\begin{equation}
 \hat{V}_{\mu} (\vec{p},\vec{k}) =
 \frac{ \langle B(\vec{p})| V_{\mu}^{\dag}|
                       \pi(\vec{k})\rangle_{\rsub{latt}} }
      { \sqrt{2 E_B(\vec{p})} \sqrt{2 E_{\pi}(\vec{k})} },
\label{eq:V_hat}
\end{equation}
because it is defined only through the residue of 
the two- and three-point correlation functions without
the knowledge how one defines the meson energies.
Since there are uncertainties in the light-light and
heavy-light meson dispersion relations, it is better to
deal with the quantity which is free from the ambiguity.
Moreover, $\hat{V}_{\mu}$ is the quantity which has the
infinite mass limit
in the heavy quark effective theory.
When the perturbative correction is incorporated,
$\Theta(m_B/m_B^{\rsub{(phys)}})$ given by
Eq.~(\ref{eq:Theta}) is multiplied to $\hat{V}_{\mu}$.
Therefore $\hat{V}_{\mu}$ is suitable quantity to study
the $1/m_Q$ dependence.

For the spatial components of $\hat{V}_{\mu}$, we also
define the scalar products
\begin{equation}
 \hat{U}_p(\vec{p},\vec{k}) =
 \frac{\vec{p}\cdot\hat{\vec{V}}(\vec{p},\vec{k})}{\vec{p}^2},
 \mbox{\hspace{1.6cm}}
\hat{U}_k(\vec{p},\vec{k})   =
 \frac{ \vec{k}\cdot\hat{\vec{V}}(\vec{p},\vec{k}) }{\vec{k}^2 }.
\label{eq:U_hat}
\end{equation}
In Table \ref{tab:V} we list the values of $\hat{V}_4$,
$\hat{U}_p$, and $\hat{U}_k$ for all momentum configurations
$(\vec{p},\vec{k})$ at $m_Q=2.6$ and $\kappa=0.1570$.
In this table, we also list the values of $q^2$ determined
with the tree
level mass relation (\ref{eq:B_mass}) for the $B$ meson.

We have investigated the $n$-dependence of $\hat{V}_{\mu}$
at $m_Q=2.1$ with $n=1$ and $2$ and at $m_Q=1.2$ with $n=2$ and $3$,
using the first 60 configurations on which $(m_Q,n)=(2.1,2)$ and
$(1.2,3)$ data are measured\footnote{
 We note that $n$-dependence should be studied on the same
 configurations.  In some of the figures, there appear large
 deviations for the data with different $n$ but the same $m_Q$.
 However, in these graphs only the results for $(m_Q,n)=(2.1,2),(1.2,3)$ 
 are obtained from the first 60 configurations and the results
 for the other combinations of $(m_Q,n)$ are obtained from the entire
 120 configurations. It seems that these large deviations seem to arise
 from the statistical fluctuation caused by the remaining
 60 configurations for which there is no the data with 
 $(m_Q,n)=(2.1,2),(1.2,3)$ .}.
For both of the heavy quark masses we observed small
dependence on $n$, which is at most 1\%, 8\% and 2\% for 
$\hat{V}_4$, $\hat{U}_p$ and $\hat{U}_k$ respectively,
and smaller than their statistical error.
In the present work, therefore, we regard them to be sufficiently
small to estimate the size of the renormalization effect 
in the manner described in Section~\ref{subsec:Pcorr}.

\subsection{Form factors}

To convert $\hat{V}_4$, $\hat{U}_p$, and $\hat{U}_k$ to
the form factors, we need to assume certain dispersion
relations for $E_B(\vec{p})$ and $E_{\pi}(\vec{k})$.
One method is to use the values obtained from the 
dispersion relation measured in the simulation.
This, however, suffers from the large statistical error for
the finite spatial momenta.
Alternatively, we adopt the following relativistic
dispersion relations for both the $B$ meson and the pion.
\begin{equation}
 E_{B}(\vec{p})   = \sqrt{m_{B}^2   + \vec{p}^2},
 \mbox{\hspace{1.6cm}}
 E_{\pi}(\vec{k}) = \sqrt{m_{\pi}^2 + \vec{k}^2},
\label{eq:DR_rel}
\end{equation}
where the measured rest mass is used for $m_{\pi}$
and $m_{B}$.
These relations are almost satisfied as shown
in Figs. \ref{fig:DR_pi} and \ref{fig:DR_B} for
light-light and heavy-light mesons, respectively.

Using the relations Eq. (\ref{eq:DR_rel}), the form factors
are easily constructed from $\hat{V}_{\mu}$.
First, we calculate $f^0(q^2)$ with
\begin{equation}
 f^0 (q^2) 
  = \frac{ \sqrt{2 E_B(\vec{p})}\sqrt{2E_{\pi}(\vec{k})} }
       {m_B^{\,2}-m_{\pi}^{\,2}} \,\, q^{\mu} \, \hat{V}_{\mu},
\end{equation}
and $f^+(q^2)$ is similarly obtained from 
$(p+k)^{\mu}\hat{V}_{\mu}$
substituting the value of $f^0$ determined above.

For $\vec{p}\neq 0$ and $\vec{k}\neq 0$,
$f^0$ and $f^+$ are not uniquely determined from 
$\hat{V}_4$, $\hat{U}_p$, and $\hat{U}_k$.
In this case there is an additional relation among
$\hat{V}_{\mu}$'s,
which should be satisfied when the Lorentz symmetry is
restored.
For $\vec{p}\perp \vec{k}$ this relation reads
\begin{equation}
 E_B(\vec{p})\hat{U}_p + E_{\pi}(\vec{k})\hat{U}_k =
 \hat{V}_4.
\label{eq:V4_condition}
\end{equation}
We examine this condition for $i_q=6, 9, 14$ and $16$ 
($i_q$ is referred in Table \ref{tab:momset}).
Figure \ref{fig:V4condition} compares LHS and RHS of 
Eq. (\ref{eq:V4_condition}) at $\kappa=0.1570$ for $i_q$=6,
with the tree level dispersion relation for $E_{B}$.
This figure exhibits a difference of about 15\%.
In other cases of $i_q$, similar amount of the discrepancy
is observed.
The size of this systematic effect is consistent with
the naive expectation for $O(a)$ error.

\subsection{Chiral extrapolation}

To obtain the form factors at the physical pion and $B$
meson masses, it is necessary to extrapolate the results
to the chiral limit.
There is, however, still a subtlety in the chiral
extrapolation, because the light quark mass dependence of
the matrix elements or the form factors are not well understood.
In principle, the chiral limit of the matrix elements or 
the form factors must be taken using the result of the
chiral effective theory as a guide for its functional form.
For the $B \rightarrow \pi$ semileptonic decay the heavy
meson effective theory with chiral Lagrangian gives such
an example \cite{Georgi91,BD-W92,KK94}.

At least the heavy meson effective theories tell us that the
matrix elements or the form factors depend on $v\cdot k$,
where $v^{\mu}$ is the 4-velocity of the $B$ meson.
At the zero pion momentum, the quantity $v \cdot k$ could
potentially give linear dependence in $m_{\pi}$, which
could result in a $\sqrt{m_q}$ dependence.
The zero recoil limit in the heavy meson effective
theory gives the following relations
for the matrix element and the form factor:
\begin{equation}
  \langle\pi(\vec{k}=0)|V_4|B(\vec{p}=0)\rangle
  = (m_B+m_{\pi}) f^0(q_{\rsub{max}}^2)
  = m_B \frac{f_B}{f_{\pi}},
\label{eq:HMET}
\end{equation}
Assuming the linear dependence of $f_B$, $f_{\pi}$,
and $m_B$ on $m_q$, at least in the zero recoil limit
the matrix element should have linear dependence on $m_q$.
In the following analysis, we take the chiral limit
of the matrix elements assuming the linear dependence
on $m_q$ in any case of $(\vec{p},\vec{k})$,
although there is no proof.

Figure \ref{fig:Chiral_ext} shows the chiral extrapolation
of the matrix element with the form 
\begin{equation}
  \langle\pi(\vec{k})|V_{\mu}|B(\vec{p})\rangle = a_V + b_V m_q,
\label{eq:chiral}
\end{equation}
where $m_q=1/2\kappa-1/2\kappa_c$.
The data itself do not show any sign of nonlinear behavior
at least around the strange quark mass.
The form factors $f^+(q^2)$ and $f^0(q^2)$ at the physical
pion mass are extracted after extrapolating
the matrix elements to the chiral limit using Eq.(\ref{eq:chiral}).

\section{PHYSICAL IMPLICATIONS}
\label{sec:PHYSICAL_IMPLICATIONS}

In this section we discuss the physical implications of our
results, which include the $1/m_B$ dependence of the 
$B\rightarrow\pi$ matrix elements and the $q^2$ dependence
of the form factors.
The prediction from the soft pion theorem is compared with
our data.

\subsection{$1/m_B$ dependence}

The heavy quark effective theory predicts that the
properly normalized $B\rightarrow\pi$ matrix element has a
static limit, hence it can be described by an expansion in the 
inverse heavy meson mass $1/m_B$ whose leading order is a
function of the heavy meson velocity $v_{\mu}=p_{\mu}/m_B$,
\begin{equation}
  \frac{\langle\pi(\vec{k})|V^{\dag}_{\mu}|B(\vec{p})\rangle}
          {\sqrt{m_{\pi}m_B}}
  = \theta_1(v\cdot k) v_{\mu}  
  + \theta_2(v\cdot k) \frac{k_{\mu}}{v\cdot k}.
\label{eq:matrix_el}
\end{equation}
Similar arguments for the heavy-light decay constant 
suggested that the quantity $f_B\sqrt{m_B}$ has the static
limit while numerical simulations have shown that 
the $1/m_B$ correction is very large.
On the other hand, the $1/m_B$ dependence of the form factors
have been studied only in the $D$ meson region
\cite{APE95,UKQCD95,Wuppertal96}.
Therefore it is important to study the $1/m_B$ dependence of
the matrix elements at fixed values of $v\cdot k$.

Except for $\vec{p}=0$, fixing $\vec{p}$ is not quite 
identical to fixing $v\cdot k$,
since the velocity $v_{\mu}$ changes depending on the heavy
meson mass.
Thus it is awkward to use the matrix elements with nonzero
$\vec{p}$.
In the special case of $\vec{p}=0$, LHS of Eq. 
(\ref{eq:matrix_el}) is nothing but the matrix elements
$\hat{V}_4$, $\hat{U}_p$ and $\hat{U}_k$,
defined in Eqs. (\ref{eq:V_hat}), and (\ref{eq:U_hat}),
multiplied by the $m_B$ independent factor.

In the following analysis, we confine ourselves to
examine the following quantities for the sake of simplicity:
\begin{eqnarray}
 \hat{V}_4(\vec{p}=0,\vec{k}) 
  &=&  \hat{V}_4^{(0)}  \left( 1 + 
   \frac{c_4^{(1)}}{m_B} + \frac{c_4^{(2)}}{m_B^{\,2}} 
                                      + \cdots \right), \\
 \hat{U}_k(\vec{p}=0,\vec{k}) 
  &=&  \hat{U}_k^{(0)}\left( 1 + 
   \frac{c_k^{(1)}}{m_B} + \frac{c_k^{(2)}}{m_B^{\,2}} 
                                     + \cdots \right), \\
 \hat{U}_p(\vec{p}=0,\vec{k}) 
  &\equiv& \mathop{\mbox{lim}}_{\vec{p}^2 \rightarrow 0}
            \hat{U}_p(\vec{p},\vec{k}) \\
  &=&  \frac{1}{m_B} \hat{U}_p^{\prime (0)} \left( 1 + 
  \frac{c_p^{(1)}}{m_B} + \frac{c_p^{(2)}}{m_B^{\,2}}
     + \cdots \right),
\end{eqnarray}
for which we explicitly show the form of the $1/m_B$
expansion.
All of the coefficients in these expansions are a function
of $\vec{k}$.

In Figs. \ref{fig:V4_kdep} and \ref{fig:Vk_kdep} we show
the $1/m_B$ dependence of $\hat{V}_4$ and $\hat{U}_k$,
respectively, at $\kappa=0.1570$. 
The $1/m_B$ correction is not significant for these
quantities and almost negligible around the $B$ meson mass.
This result exhibits a sharp contrast to the mass
dependence of the heavy-light decay constant 
$f_B\sqrt{m_B}$, for which the large $1/m_B$ correction to
the static limit is observed.
Results of the linear and quadratic fit in $1/m_B$ are
listed in Table \ref{tab:V4_fit} for $\hat{V}_4$ and
in Table \ref{tab:Vk_fit} for $\hat{U}_k$. 

We note here that $\chi^2$/dof are less than unity for 
most cases of $V_4$, $U_k$, and also $U_p$, which will be
mentioned in the next paragraph,
though they do not exactly judge the goodness of the fits
for such data, which are correlated for different $m_Q$.

In order to do the same discussion for $\hat{U}_p$,
which is defined in the $\vec{p}^2\rightarrow 0$ limit,
we extrapolate the finite $\vec{p}$ results to the vanishing
$\vec{p}$ point as shown in Fig. \ref{fig:Vp_ext}.
There is little $\vec{p}^2$ dependence observed and we
employ a linear extrapolation in $\vec{p}^2$.
In Fig. \ref{fig:Vp_kdep} we plot $m_B \hat{U}_p$ as
a function of $1/m_B$ at $\kappa=0.1570$. 
In contrary to the other matrix elements we observe a
sizable $1/m_B$ dependence. 
Table \ref{tab:Vp_fit} summarizes the results of linear and
quadratic fit of $m_B \hat{U}_p$.

Here we briefly discuss the effect of one-loop correction
to these quantities.
Figure \ref{fig:V_renorm} shows the renormalized values of
$\hat{V}_4(i_q=1)$, $\hat{U}_k(i_q=2)$, and
$m_B\hat{U}_p(i_q=1)$ at $\kappa=0.1570$. 
As mentioned at the end of Section \ref{sec:LATTICE_NRQCD},
the leading logarithmic factor Eq.~(\ref{eq:Theta}) is
multiplied to $\hat{V}_{\mu}$.
We also list the results of linear fits of them in Table
\ref{tab:V_renorm}.
As we discussed previously, the $1/m_Q$ dependence of
the one-loop coefficient is significant only for $V_4$
and almost negligible for $V_i$. 
As a result, the $1/m_B$ dependence of $\hat{V}_4$ is largely
affected by the renormalization effect, and it even changes
the sign of the slope in $1/m_B$. 
The $1/m_B$ dependence of $\hat{V}_4$ is still mild
after the renormalization effect is included.
For $\hat{U}_k$ and $m_B \hat{U}_p$ the $1/m_B$ dependence is
not affected by the one-loop correction,
while their amplitudes decrease by at most 30\%.

\subsection{$q^2$-dependence of the form factors}

First we study for which $q^2$ region our present 
statistics allow us to compute the form factors 
with reasonable statistical errors.
The $q^2$ dependence of the form factors $f^+$ and $f^0$
are shown in Figs. \ref{fig:FF_02} and \ref{fig:FF_04}
at $m_Q$=2.6 and 1.5, respectively. 
We find that for $\kappa=0.1570$($m_q \sim 2 m_s$),
the range of $q^2$ in which the form factors have good
signal covers almost the entire kinematic region for $D$
meson and one third of the kinematic region for $B$ meson.
For $\kappa=0.1600$ ($m_q \sim m_s$), the signal becomes 
much noisier, but still the form factors have marginally
good signal for half and one fourth of the kinematic region
for $D$ meson and $B$ meson, respectively.
Although our present results are very noisy after the
chiral extrapolation, this will be improved by future high
statistics studies.
This is encouraging in view of the fact that the future 
$B$ Factories can produce $10^8$ $B$-$\overline{B}$ pairs
and the branching fraction of $B \rightarrow \pi l \nu$ from
CLEO is $(1.8 \pm 0.4 \pm 0.3 \pm 0.2) \times 10^{-4}$
\cite{CLEO96}.
It is reasonable to expect that there is a possibility
of observing $B \rightarrow \pi l \nu$ events in the $q^2$
regime which the present lattice calculation can cope with.

Secondly we study the $q^{2}$ dependence to see 
whether the contribution from  the $B^*$ resonance to
the form factor can actually be observed in the simulation
data.
At the chiral limit, unfortunately, the results are
too noisy to discuss their $q^2$ dependence,
therefore we use the finite mass results only in the 
following analysis of the $q^2$ dependence.
As shown in Figs. \ref{fig:FF_02} and \ref{fig:FF_04},
the lattice results are available
only in the large $q^2$ region, at which the recoil
momentum of pion is small enough.
Therefore it is justified to express the functional form of
the form factors by an expansion around the zero recoil
limit.
For this purpose we use the inverse form factors
$1/f^+(q^2)$ and $1/f^0(q^2)$:
\begin{equation}
  1/f(q^2) = 1/f(q^2_{\rsub{max}})
   + c_1 (q^2_{\rsub{max}}-q^2) 
   + c_2 (q^2_{\rsub{max}}-q^2)^2.
\label{eq:ffit}
\end{equation}
Figure \ref{fig:FFr_02} shows the inverse form factors at
$m_Q=2.6$ as well as their fitted functions with this form.
The numerical results of the fit with and without the
condition $c_2=0$ are given in Table \ref{tab:FF_fit} for
$m_Q=2.6$, $1.5$, and $0.9$. 

The pole dominance model corresponds to a special case
$c_2=0$, which seems to describe the data very well as
shown in Fig. \ref{fig:FFr_02}. 
The mass of the intermediate state is given by
$m_{\rsub{pole}}^2=q_{\rsub{max}}^2
+1/(c_1 f^+(q_{\rsub{max}}^2))$,
which corresponds to the vector ($B^*$) meson mass
in the pole dominance model.
Precisely speaking, the more consistent analysis is to
impose the condition $m_{\rsub{pole}}=m_{B^*}$
for the fit by Eq. (\ref{eq:ffit}). 
This constrained fit is shown with
the long dashed line in Fig. \ref{fig:FFr_02}.
It is found that now the fit do not quite agree with
the data, but the deviation is about 10 \%.  
  
In Fig. \ref{fig:pole_mass} we also compare
$m_{\rsub{pole}}$ and
the measured vector meson mass as a function of $m_B$. 
Again we find that there is a discrepancy between
$m_{\rsub{pole}}$
from the unconstrained fit and the measured $m_{B^*}$,
which is around few hundred MeV.
Nevertheless, it is remarkable that the deviation remains 
the same order and the mass dependence of $m_{\rsub{pole}}$
has the same trend with $m_{B^*}$.
We have not yet understood whether the above discrepancies
can be explained from the remaining systematic errors such
as the discretization error.
But at least qualitatively judging from the size of the
uncertainty in our calculation, our data is not
inconsistent with the picture that there is a sizable
contribution from the $B^*$ pole to the form factor
$f^{+}$ near $q^2_{\rsub{max}}$.

So far the discussion have been based on the tree level
study.
Let us now study how one-loop renormalization changes
the form factors.
Because the one-loop correction is different for $V_4$ and
$V_i$, the shape of the form factors may change significantly.
Figure \ref{fig:FF02renorm} shows the form factors
for $m_Q=2.6$ and $\kappa=0.1570$ with renormalization
factors. 
The leading logarithmic factor Eq.(\ref{eq:Theta})
is not multiplied in the present case.
We find that the renormalized $f^+$ has stronger $q^2$
dependence than that of at the tree level,
while $f^0$ receives only a small change.
The renormalization makes the $B^*$ pole fit even worse.
In fact, the deviation of the constrained fit 
from our renormalized $f^+$ data is as large as 25 \%
near $q^2_{\rsub{max}}$.
This is still within the typical size of
$O(a)$ errors.
It is very important to perform the analysis with larger
$\beta$.

\subsection{Soft pion theorem}

Applying the soft pion theorem to the $B\rightarrow\pi$
matrix element, $f^0(q_{\rsub{max}}^2)$ is related to
the $B$ meson decay constant \cite{BD-W92,KK94,BLNN94}
\begin{equation}
  f^0(q^2_{\rsub{max}}) = f_B/f_{\pi}.
\end{equation}
in the massless pion limit.
This relation is examined in Fig. \ref{fig:SPT}.
For the values of $f_B$, we refer our work on 
$f_B$\cite{Hiroshima97}, which is obtained with an
evolution equation of a slightly different form from that
of the present work.
We observe a large discrepancy between $f^0$ and the decay
constant both for the $1/m_B$ dependence and for the value
itself.
$f_B$ increases rapidly toward heavier heavy quark masses,
while $f^0(q^2_{\rsub{max}})$ almost stays constant.

The discrepancy still remains significant when the
renormalization effect is incorporated.
In evaluating the renormalized values of $f_B$,
we use one-loop perturbative coefficient obtained in the same
manner as in Section~\ref{subsec:Pcorr} \cite{Ishikawa97}.
The leading logarithmic factor Eq.(\ref{eq:Theta}) is 
multiplied to both $f^0(q^2_{\rsub{max}})$ and $f_B$. 

One may argue that the observed discrepancy can be
explained by the uncertainty in the extrapolation procedure.
To study this possibility, we compare $f^0(q^2_{\rsub{max}})$
and $f_B/f_{\pi}$ also in finite light quark mass cases,
in the light of the heavy meson effective theory which
implies the relation (\ref{eq:HMET}).
They are compared in Fig. \ref{fig:SPT_ext} as a function
of $1/\kappa$.
The difference between them are remarkable even for
finite light quark mass cases.

The reason why these differences occur is not clear. 
Since our present results suffer from various systematic
uncertainties, as described in the next section, 
further study with better control of systematic errors
is necessary to clarify the origin of
the problem.

\section{SYSTEMATIC ERRORS}
\label{sec:SYSTEMATIC_ERRORS}

In this section, we qualitatively discuss on the systematic
uncertainties associated with the lattice regularization.
The following is a list of the main sources of systematic
errors:
\begin{itemize}
\item $O(a)$ errors:
  The characteristic size of $O(a\Lambda_{QCD})$ error
  arising from the unimproved Wilson quark action at
  $\beta=5.8$ is 20--30\%.
  This effect is large enough to explain the discrepancy
   between
  $E_B(\vec{p})\hat{U}_p+E_{\pi}(\vec{k})\hat{U}_k$
  and $\hat{V}_4$, mentioned in Section 
  \ref{sec:SIMULATION_DETAILS}.
  Use of the $O(a)$-improved Clover action for the light
  quark will reduce this error to the level of 5 \%.
\item $O((ap)^2)$ error:
  The systems with finite momentum may suffer from the
  discretization errors more seriously than that at the
  zero recoil point.
  The analytic estimate of the momentum dependent error
  \cite{Simone96} shows that the effect is about 20 \% 
  at $|\vec{p}|\sim 1$ GeV even one uses the 
  $O(a)$-improved current.
\item Perturbative corrections:
  The one-loop correction could become significant
  especially for small $\beta$ values.
  Strictly speaking, our calculation does not treat the
  one-loop effects correctly, because the stabilizing
  parameter $n$ does not have correct values.
  This problem must be removed in the future studies.
  In estimating the one-loop corrections,
  we did not include the effect of the operator mixing,
  which was reported to be significant in the case of 
  $f_B$ \cite{Shigemitsu97b}.
  This effect also should be included to obtain reliable
  results. 
\item $O(1/m_Q^{\,2})$ effects:
   We described the heavy quark with the NRQCD action
   including the order $1/m_Q$ terms.
   Further precise calculations may need to include
   $O(1/m_Q^{\,2})$ corrections, although the effect was
   shown to be
   small\cite{Hiroshima97} for $f_B$.
\end{itemize}
The finite volume effect may also be important.

Since the all above systematic errors can be large,
there is no advantage of giving quantitative estimates
of each error at this stage.
The use of the $O(a)$-improved (clover) action for light
quark, as well as the simulation at higher $\beta$ values
will reduce most of the above systematic errors.
The simulation with dynamical quarks is also of great
importance for reliable predictions of the weak matrix
elements.

\section{CONCLUSION}
\label{sec:CONCLUSION}

In this paper, we present the results of the study of
$B\rightarrow\pi$ form factors using NRQCD to describe
the heavy quark with the Wilson light quark. 
Clear signal is observed for the matrix element in a wide
range of heavy quark mass containing the physical $b$-quark
mass. 
They are extrapolated to the chiral limit, although the result
is so noisy for quantitative conclusion.

The $1/m_B$ dependence of the matrix elements are studied and
it is clarified that the temporal component and the part
of the spatial component proportional to the pion momentum
have fairly small dependencies on $m_Q$.
On the other hand, the part of the spatial component
proportional to the $B$ momentum has a significant
$O(1/m_B)$ correction.

The $q^2$ dependence of the form factors in the finite
light quark masses are studied.
We find that the $q^2$ dependence of the form factor $f^+(q^2)$
near $q_{\rsub{max}}^2$ becomes much stronger for larger heavy quark mass.
Model independent fit of $1/f^{+}(q^2)$ near $q_{\rsub{max}}^2$ 
shows that the tree level results are consistent with the
pole behavior for large $q^2$ range, and the difference of 
fitted pole mass and the measured $m_{B^{*}}$ is 
around few hundred MeV for all the heavy quark masses.

The values of $f^0$ at the zero recoil point are compared
with the prediction of the soft pion theorem, and the
significant discrepancy is observed.

The size of the renormalization corrections are estimated 
by the one-loop perturbative calculation.
They almost does not affect their $1/m_Q$ dependence,
but decrease $V_i$ much more than $V_4$, which drastically
change the shape of $f^+$.
Our present result suffers from large systematic
uncertainties, and the most important one is $O(a)$ error.
It is very important to study at higher $\beta$ with
improved actions.

\section*{ACKNOWLEDGMENT}

Numerical simulations were carried out on Intel Paragon XP/S
at INSAM (Institute for Numerical Simulations and Applied
Mathematics) in Hiroshima University.
We are grateful S. Hioki and O. Miyamura for 
kind advice.
We thank members of JLQCD collaboration for useful
discussions. 
H.M. would like to thank the Japan Society for the
Promotion of Science for Young Scientists for 
financial support.
S.H. is supported by Ministry of Education, Science and
Culture under grant number 09740226.

\clearpage

\begin{table}[p]
\begin{center}
\begin{tabular}{ccccc}
\makebox[2.4cm]{ $(m_Q\,,\,n)$ } &
\makebox[2.4cm]{ $A$ } & 
\makebox[2.4cm]{ $B$ } & 
\makebox[2.4cm]{ $C_{V_4}$ } & 
\makebox[2.4cm]{ $C_{V_i}$ } \\ 
\hline
(5.0\,,\,1) & 0.0759 & 0.0124(4) &    0.0210(11)& $-$0.0790(10) \\
(2.6\,,\,2) & 0.0668 & 0.0353(3) &    0.0004(9) & $-$0.0780(7)  \\
(2.1\,,\,2) & 0.0623 & 0.0449(3) & $-$0.0068(9) & $-$0.0757(7)  \\
(1.5\,,\,3) & 0.0528 & 0.0623(2) & $-$0.0192(8) & $-$0.0734(6)  \\
(1.2\,,\,3) & 0.0446 & 0.0757(1) & $-$0.0283(8) & $-$0.0707(6)  \\
(0.9\,,\,6) & 0.0309 & 0.0933(1) & $-$0.0428(8) & $-$0.0687(5)  \\
\end{tabular}
\end{center}
\caption{ 
The tadpole improved one-loop coefficients for the perturbative
corrections $E_0$, $Z_m$, $Z_{V_4}$, and $Z_{V_i}$. 
Quoted errors represent the numerical uncertainties in the
evaluation of loop integrals.
The uncertainty of $A$ is less than $10^{-4}$.   }
\label{tab:P_corr}
\end{table}

\begin{table}[p]
\begin{center}
\begin{tabular}{ccccc}
\makebox[2.0cm]{  } &
\makebox[2.8cm]{ $\kappa=0.1570$ } & 
\makebox[2.8cm]{ $0.1585$ } & 
\makebox[2.8cm]{ $0.1600$ } & 
\makebox[2.8cm]{ $\kappa_c$ } \\ 
\hline
 $m_{\pi}$  & 0.5677(30) & 0.4933(33) & 0.4118(37) &     -      \\
 $m_{\rho}$ & 0.6747(54) & 0.6214(72) & 0.567(11)  & 0.448 (17) \\
 $f_{\pi}$  & 0.1496(46) & 0.1380(49) & 0.1270(53) & 0.1019(64) \\
\end{tabular}
\end{center}
\caption{The values of $m_{\pi}$, $m_{\rho}$, 
and pion decay constant without renormalization. 
Fitting range is $t=14-24$. }
\label{tab:Spc_pi}
\end{table}

\begin{table}[p]
\begin{center}
\begin{tabular}{clllcccc}
\makebox[0.8cm]{ $i_q$ } &
\makebox[0.8cm]{ $\vec{p}^{\,2}$ \ \ \ } &
\makebox[1.2cm]{ $\vec{k}^{\,2}$ \ \ \ } &
\makebox[0.8cm]{ $\vec{q}^{\,2}$ \ \ \ } &
\makebox[2.2cm]{ $\vec{p}$  } &
\makebox[2.2cm]{ $\vec{k}$  } &
\makebox[2.4cm]{ $-\vec{q}=\vec{k}-\vec{p}$ } &
\makebox[1.2cm]{ $\sharp$ $(\vec{p}, \vec{k})$ } \\
\hline 
 1 & 0 & 0   & 0 & ( 0,  0,   0 )    
                  & ( 0, 0, 0 ) & (  0,  0,  0 ) &  1   \\ 
 2 &   & 1   & 1 &           
                  & ( 0, 0, 1 ) & (  0,  0,  1 ) &  6  \\
 3 &   & 2   & 2 &   
                  & ( 0, 1, 1 ) & (  0,  1,  1 ) & 12  \\
 4 &   & 3   & 3 & 
                  & ( 1, 1, 1 ) & (  1,  1,  1 ) &  8  \\
\hline
 5 & 1 & 0   & 1 & ( 0,  0,  1 ) 
               & ( 0, 0, 0 ) & (  0,  0, $-$1 ) &  6   \\
 6 &   & 1($\perp$) & 2 & ( 0,  1,  0 ) 
               & ( 0, 0, 1 ) & (  0, $-$1,  1 ) & 24   \\
 7 &   & 1($\uparrow\!\uparrow$) & 0 & ( 0,  0,  1 ) 
                 & ( 0, 0, 1 ) & (  0,  0,  0 ) &  6   \\
 8 &   & 1($\uparrow\!\downarrow$) & 4 & ( 0,  0, $-$1 ) 
                 & ( 0, 0, 1 ) & (  0,  0,  2 ) &  2   \\
 9 &   & 2($\perp$) & 3 & ( 1,  0,  0 ) 
               & ( 0, 1, 1 ) & ( $-$1,  1,  1 ) & 24   \\
10 &   & 2   & 1 & ( 0,  0,  1 ) 
                 & ( 0, 1, 1 ) & (  0,  1,  0 ) & 24   \\
11 &   & 3   & 2 & ( 0,  0,  1 ) 
                 & ( 1, 1, 1 ) & (  1,  1,  0 ) & 24   \\
12 &   & 3   & 6 & ( 0,  0, $-$1 ) 
                 & ( 1, 1, 1 ) & (  1,  1,  2 ) &  8   \\
\hline
13 & 2 & 0   & 2 & ( 0,  1,  1 ) 
             & ( 0, 0, 0 ) & (  0, $-$1, $-$1 ) & 12   \\
14 &   & 1($\perp$) & 3 & ( 1,  1,  0 ) 
             & ( 0, 0, 1 ) & ( $-$1, $-$1,  1 ) & 24   \\
15 &   & 1   & 1 & ( 0,  1,  1 ) 
               & ( 0, 0, 1 ) & (  0, $-$1,  0 ) & 24   \\
16 &   & 2($\perp$) & 4 & ( 0,  1, $-$1 ) 
                 & ( 0, 1, 1 ) & (  0,  0,  2 ) &  4   \\
17 &   & 2($\uparrow\!\uparrow$) & 0 & ( 0,  1,  1 ) 
                 & ( 0, 1, 1 ) & (  0,  0,  0 ) & 12   \\
18 &   & 2   & 2 & ( 1,  1,  0 ) 
               & ( 0, 1, 1 ) & ( $-$1,  0,  1 ) & 48   \\
19 &   & 2   & 6 & ( 1, $-$1,  0 ) 
               & ( 0, 1, 1 ) & ( $-$1,  2,  1 ) & 16   \\
\hline
20 & 3 & 0 & 3 & ( 1, 1, 1 ) 
           & ( 0, 0, 0 ) & ( $-$1, $-$1, $-$1 ) &  8   \\
\end{tabular}
\end{center}
\caption{
The momentum combinations $(\vec{p},\vec{k})$ used in 
the simulation. 
In this table, the values of $\vec{p}$, $\vec{k}$, 
and $\vec{q}$ are expressed in the unit of  $2\pi/16$.
The set which is equivalent with another under the 
lattice rotational 
symmetry is identified with the same $i_q$-number, and a
representative is shown in the fifth through seventh columns.
The last column shows the numbers of equivalent combinations.
The symbols in the third column denote the direction 
of $\vec{k}$ against $\vec{p}$ as follows:
$\perp$: orthogonal,  $\uparrow\!\uparrow$: parallel, 
$\uparrow\!\downarrow$: anti-parallel, and oblique for
others.
The set $i_q=12$  gives the minimum $q^2$ value among 
the sets in this table.  }
\label{tab:momset}
\end{table}

\begin{table}[p]
\begin{center}

Pseudoscalar meson binding energy: $E_{\bar{q}Q}(\vec{p}=0)$

\begin{tabular}{ccccc}
\makebox[2.0cm]{ $(m_Q\,,\,n)$ } &
\makebox[2.8cm]{ $\kappa=0.1570$ } & 
\makebox[2.8cm]{ $0.1585$ } & 
\makebox[2.8cm]{ $0.1600$ } & 
\makebox[2.8cm]{ $\kappa_c$ } \\ 
\hline
(5.0\,,\,1)& 0.6304(69)& 0.6084(83)& 0.585 (11)& 0.535 (15) \\
(2.6\,,\,1)& 0.6268(48)& 0.6041(56)& 0.5809(71)& 0.530 (10) \\
(2.1\,,\,1)& 0.6247(45)& 0.6014(52)& 0.5777(65)& 0.5260(91) \\
(2.1\,,\,2)& 0.6279(53)& 0.6056(62)& 0.5834(80)& 0.534 (11) \\
(1.5\,,\,2)& 0.6180(42)& 0.5940(48)& 0.5696(59)& 0.5162(81) \\
(1.2\,,\,2)& 0.6135(40)& 0.5889(46)& 0.5640(56)& 0.5095(75) \\
(1.2\,,\,3)& 0.6142(51)& 0.5899(56)& 0.5655(68)& 0.5117(92) \\
(0.9\,,\,2)& 0.6058(39)& 0.5805(43)& 0.5551(51)& 0.4991(68) \\
\end{tabular}
\vspace{0.2cm}

Vector meson binding energy: $E_{\bar{q}Q^*}(\vec{p}=0)$

\begin{tabular}{ccccc}
\makebox[2.0cm]{ $(m_Q\,,\,n)$ } &
\makebox[2.8cm]{ $\kappa=0.1570$ } & 
\makebox[2.8cm]{ $0.1585$ } & 
\makebox[2.8cm]{ $0.1600$ } & 
\makebox[2.8cm]{ $\kappa_c$ } \\ 
\hline
 (5.0\,,\,1) & 0.649  (12) & 0.628  (14) & 0.604  (19) & 0.555 (27) \\
 (2.6\,,\,1) & 0.6502 (62) & 0.6287 (76) & 0.6065 (99) & 0.559 (14) \\
 (2.1\,,\,1) & 0.6501 (56) & 0.6279 (68) & 0.6047 (88) & 0.555 (13) \\
 (1.5\,,\,2) & 0.6488 (52) & 0.6257 (61) & 0.6014 (79) & 0.550 (11) \\
 (1.2\,,\,2) & 0.6484 (51) & 0.6249 (59) & 0.6002 (76) & 0.547 (11) \\
 (0.9\,,\,2) & 0.6470 (50) & 0.6231 (57) & 0.5982 (73) & 0.545 (10) \\
\end{tabular}
\end{center}
\caption{ The binding energy of the pseudoscalar and vector
heavy-light mesons. 
The single exponential fit is applied with the fitting range $t=16-24$.
For the pseudoscalar we average the values obtained from the
local-local and the smeared-local correlation functions.
For the vector mesons we use the local-local only, and there is no
data available for $(m_Q,n)=(2.1,2)$ and $(1.2,3)$. }
\label{tab:Spc_B}
\end{table}

\begin{table}[p]
\begin{center}
\begin{tabular}{cccccc}
\makebox[18mm]{ $i_q$ } & 
\makebox[28mm]{ $q^2$ } & 
\makebox[28mm]{ $\hat{V}_4$ } & 
\makebox[28mm]{ $\hat{U}_p$ } & 
\makebox[28mm]{ $\hat{U}_k$ } \\ 
\hline
  1 & 7.071 (20) & 1.014 (34) &       -       &      -     \\
  2 & 6.280 (19) & 0.844 (26) &       -       & 0.878 (41) \\
  3 & 5.609 (19) & 0.754 (50) &       -       & 0.695 (61) \\
  4 & 5.017 (18) & 0.612 (87) &       -       & 0.57  (10) \\
\hline
  5 & 7.044 (20) & 0.999 (36) &   0.0475(28) &      -     \\
  6 & 6.247 (19) & 0.832 (28) &   0.0366(47) & 0.860 (41) \\
  7 & 6.555 (19) & 0.930 (30) &   1.009 (46) & 1.009 (46) \\
  8 & 5.938 (19) & 0.750 (34) &$-$0.702 (48) & 0.702 (48) \\
  9 & 5.571 (19) & 0.742 (49) &   0.040 (12) & 0.674 (59) \\
 10 & 5.880 (19) & 0.827 (55) &   0.790 (68) & 0.767 (66) \\
 11 & 5.283 (18) & 0.66  (10) &   0.65  (12) & 0.63  (11) \\
 12 & 4.666 (18) & 0.544 (68) &$-$0.39  (12) & 0.477 (82) \\
\hline
 13 & 7.017 (20) & 0.992 (42) &   0.0467(30) &    -       \\
 14 & 6.214 (19) & 0.825 (34) &   0.0360(48) & 0.848 (45) \\
 15 & 6.523 (19) & 0.923 (38) &   0.517 (26) & 0.997 (51) \\
 16 & 5.534 (19) & 0.757 (76) &   0.052 (53) & 0.670 (82) \\
 17 & 6.151 (19) & 0.920 (67) &   0.863 (77) & 0.863 (77) \\
 18 & 5.842 (19) & 0.820 (58) &   0.412 (36) & 0.758 (68) \\
 19 & 5.225 (19) & 0.669 (52) &$-$0.266 (41) & 0.587 (61) \\
\hline
 20 & 6.990 (20) & 0.968 (58) &  0.0454 (33) &    -       \\
\end{tabular}
\end{center}
\caption{ 
$\hat{V}_4$, $\hat{U}_p$, and $\hat{U}_k$ in the lattice unit at
$m_Q=2.6$ and $\kappa=0.1570$.
$i_q$ denotes the set of momentum $(\vec{p},\vec{k})$ summarized in
Table \ref{tab:momset}.
In the evaluation of $q^2$, the $B$ meson mass is determined through
the tree level relation $m_B=m_Q+E_{\bar{q}Q}(0)$. }
\label{tab:V}
\end{table}

\begin{table}[p]
\begin{center}
\begin{tabular}{ccccccc}
 & & \multicolumn{2}{c}{linear} & \multicolumn{3}{c}{quadratic} \\
\makebox[1.2cm]{ $\kappa$ } &
\makebox[0.8cm]{ $i_q$} &
\makebox[1.2cm]{ $\hat{V}_4^{(0)}$ } & 
\makebox[2.4cm]{ $c_4^{(1)}$ } & 
\makebox[2.4cm]{ $\hat{V}_4^{(0)}$ } & 
\makebox[1.2cm]{ $c_4^{(1)}$ } & 
\makebox[2.4cm]{ $c_4^{(2)}$ } \\
\hline
$0.1570$
  & 1& 0.965(35)&   0.184(55)& 1.003(47)&$-$0.01(20)& 0.21(18)\\
  & 2& 0.826(29)&   0.080(47)& 0.851(41)&$-$0.06(17)& 0.15(17)\\
  & 3& 0.757(51)&$-$0.038(59)& 0.799(57)&$-$0.30(20)& 0.31(22)\\
  & 4& 0.624(80)&$-$0.25 (11)& 0.79 (10)&$-$1.29(36)& 1.25(42)\\
\hline
$0.1585$
  & 1& 0.982(42)&   0.165(63)& 1.016(55)&$-$0.00(23)& 0.18(21)\\
  & 2& 0.807(35)&   0.075(57)& 0.830(48)&$-$0.06(20)& 0.14(19)\\
  & 3& 0.758(76)&$-$0.071(73)& 0.830(81)&$-$0.51(26)& 0.51(29)\\
  & 4& 0.62 (12)&$-$0.40 (15)& 0.89 (19)&$-$1.83(50)& 1.75(60)\\
\hline
$0.1600$
  & 1& 1.003(53)&   0.150(76)& 1.023(66)&   0.05(27)& 0.10(25)\\
  & 2& 0.768(46)&   0.088(76)& 0.788(58)&$-$0.04(26)& 0.14(25)\\
  & 3& 0.78 (14)&$-$0.17 (10)& 0.96 (17)&$-$1.13(40)& 1.13(46)\\
  & 4& 0.70 (27)&$-$0.64 (25)& 1.22 (55)&$-$2.45(80)& 2.26(94)\\
\end{tabular}
\end{center}
\caption{ Parameters for the linear and quadratic fits of 
$\hat{V}_4(\vec{p}=0,\vec{k})$.}
\label{tab:V4_fit}
\end{table}

\begin{table}[p]
\begin{center}
\begin{tabular}{ccccccc}
 & & \multicolumn{2}{c}{linear} 
   & \multicolumn{3}{c}{quadratic} \\
\makebox[1.2cm]{ $\kappa$ } &
\makebox[0.8cm]{ $i_q$} &
\makebox[1.2cm]{ $\hat{U}_k^{(0)}$ } & 
\makebox[2.4cm]{ $c_k^{(1)}$ } & 
\makebox[2.4cm]{ $\hat{U}_k^{(0)}$ } & 
\makebox[1.2cm]{ $c_k^{(1)}$ } & 
\makebox[2.4cm]{ $c_k^{(2)}$ } \\
\hline
$0.1570$ 
 & 2& 0.945(39)&$-$0.194(44)& 0.967(47)&$-$0.30(19)&   0.13(19)\\
 & 3& 0.762(56)&$-$0.257(53)& 0.750(54)&$-$0.17(22)&$-$0.10(24)\\
 & 4& 0.655(88)&$-$0.364(91)& 0.600(81)&   0.08(43)&$-$0.54(49)\\
\hline
$0.1585$ 
 & 2& 1.004(52)&$-$0.198(50)& 1.023(58)&$-$0.28(22)&   0.10(23)\\
 & 3& 0.808(92)&$-$0.242(64)& 0.769(80)&   0.00(30)&$-$0.29(32)\\
 & 4& 0.72 (15)&$-$0.34 (14)& 0.58 (12)&   0.77(74)&$-$1.34(80)\\
\hline
$0.1600$ 
 & 2& 1.064(73)&$-$0.214(62)& 1.063(77)&$-$0.21(29)&   0.00(30)\\
 & 3& 0.92 (20)&$-$0.219(90)& 0.80 (16)&   0.47(50)&$-$0.81(53)\\
 & 4& 0.94 (37)&$-$0.23 (26)& 0.55 (23)&   3.3 (2.3)&$-$4.1 (2.4)\\
\end{tabular}
\end{center}
\caption{ Parameters for the linear and quadratic fits of 
$\hat{U}_k(\vec{p}=0,\vec{k})$.}

\label{tab:Vk_fit}
\end{table}

\begin{table}[p]
\begin{center}
\begin{tabular}{ccccccc}
 & & \multicolumn{2}{c}{linear} & \multicolumn{3}{c}{quadratic} \\
\makebox[1.2cm]{ $\kappa$ } &
\makebox[0.8cm]{ $i_q$} &
\makebox[1.2cm]{ $\hat{U}_p^{\prime (0)}$ } & 
\makebox[2.4cm]{ $c_p^{(1)}$ } & 
\makebox[2.4cm]{ $\hat{U}_p^{\prime (0)}$ } & 
\makebox[1.2cm]{ $c_p^{(1)}$ } & 
\makebox[2.4cm]{ $c_p^{(2)}$ } \\
\hline
$0.1570$ 
  & 1& 0.0887(80)& 2.61(39)& 0.0717(95)& 4.5(1.2)& $-$1.55(76) \\
  & 2& 0.089 (14)& 1.29(38)& 0.072 (13)& 2.7(1.1)& $-$1.31(88) \\
\hline
$0.1585$
  & 1& 0.0872(94)& 2.65(47)& 0.066 (11)& 5.3(1.7)& $-$2.1 (1.0) \\
  & 2& 0.093 (20)& 0.98(42)& 0.080 (18)& 2.0(1.2)& $-$1.0 (1.1) \\
\hline
$0.1600$
  & 1& 0.088 (12)& 2.72(59)& 0.059 (15)& 6.7(2.7)& $-$3.1 (1.7) \\
  & 2& 0.104 (33)& 0.67(47)& 0.097 (27)& 1.1(1.5)& $-$0.4 (1.5) \\
\end{tabular}
\end{center}
\caption{ Parameters for the linear and quadratic fits of 
$\hat{U}_p(\vec{p}=0,\vec{k})$. }
\label{tab:Vp_fit}
\end{table}

\clearpage

\begin{table}[p]
\begin{center}

$\hat{V}_4(\vec{p}=0,\vec{k}=0)$ \ ( $i_q=1$ ) \\
\begin{tabular}{ccccc}
 & \multicolumn{2}{c}{$q^*=\pi/a$} & \multicolumn{2}{c}{$q^*=1/a$} \\
\makebox[1.2cm]{ $\kappa$ } &
\makebox[2.4cm]{ $V_4^{(0)}$ } & \makebox[2.4cm]{ $c_4^{(1)}$ } & 
\makebox[2.4cm]{ $V_4^{(0)}$ } & \makebox[2.4cm]{ $c_4^{(1)}$ } \\ 
\hline
$0.1570$ &1.002(36)&0.052(55)&1.088(39)&$-$0.209(47)\\
$0.1585$ &1.019(44)&0.039(63)&1.105(46)&$-$0.216(55)\\
$0.1600$ &1.039(55)&0.030(77)&1.126(58)&$-$0.219(66)\\
\end{tabular}
\vspace{0.4cm} 

$\hat{U}_k(\vec{p}=0,|\vec{k}|=1 )$ \ ( $i_q=2$ ) \\
\begin{tabular}{ccccc}
 & \multicolumn{2}{c}{$q^*=\pi/a$} & \multicolumn{2}{c}{$q^*=1/a$} \\
\makebox[1.2cm]{ $\kappa$ } &
\makebox[2.4cm]{ $\hat{U}_k^{(0)}$ } & \makebox[2.4cm]{ $c_k^{(1)}$ } & 
\makebox[2.4cm]{ $\hat{U}_k^{(0)}$ } & \makebox[2.4cm]{ $c_k^{(1)}$ } \\ 
\hline
$0.1570$ &0.732(31)&   0.013(61)&0.609(27)&0.081(70)\\
$0.1585$ &0.778(42)&   0.005(68)&0.649(36)&0.070(78)\\
$0.1600$ &0.826(59)&$-$0.019(84)&0.689(50)&0.043(96)\\
\end{tabular}
\vspace{0.4cm} 

$\hat{U}_p(\vec{p}=0,\vec{k}=0)$ \ ( $i_q=1$ ) \\
\begin{tabular}{ccccc}
 & \multicolumn{2}{c}{$q^*=\pi/a$} & \multicolumn{2}{c}{$q^*=1/a$} \\
\makebox[1.2cm]{ $\kappa$ } &
\makebox[2.4cm]{ $\hat{U}_p^{(0)}$ } & \makebox[2.4cm]{ $c_p^{(1)}$ } & 
\makebox[2.4cm]{ $\hat{U}_p^{(0)}$ } & \makebox[2.4cm]{ $c_p^{(1)}$ } \\ 
\hline
$0.1570$ &0.0466(66)&6.3(1.2)&0.0268(58)&11.3(30)\\
$0.1585$ &0.0453(77)&6.5(1.5)&0.0256(68)&11.8(38)\\
$0.1600$ &0.045 (10)&6.7(1.9)&0.0248(87)&12.5(53)\\
\end{tabular}

\end{center}
\caption{ 
Parameters for the linear fit of the renormalized matrix elements
$\hat{V}_4(\vec{p}=0,\vec{k}=0)$, 
$\hat{U}_k(\vec{p}=0,|\vec{k}|=1)$, and 
$\hat{U}_p(\vec{p}=0,\vec{k}=0)$. }
\label{tab:V_renorm}
\end{table}

\begin{table}[p]
\begin{center}
\begin{tabular}{ccccccc}
& & \multicolumn{2}{c}{linear fit} &
    \multicolumn{3}{c}{quadratic fit} \\
\makebox[1.4cm]{ $(m_Q,n)$ } & \makebox[1.4cm]{$\kappa$}&
\makebox[1.8cm]{$f^{-1}(q^2_{\rsub{max}})$}
                                & \makebox[1.8cm]{$c_1$}& 
\makebox[1.8cm]{$f^{-1}(q^2_{\rsub{max}})$}
                                & \makebox[1.8cm]{$c_1$}& 
                                  \makebox[1.8cm]{$c_2$}\\ 
\hline
(2.6,\,\,1)
 &0.1570& 1.373(54)& 0.126(70)& 1.386(52)& 0.058(64)&   0.046(53)\\
 &      & 0.480(21)& 0.264(38)& 0.470(20)& 0.335(40)&$-$0.051(37)\\
\cline{2-7}
 &0.1585& 1.436(70)& 0.109(88)& 1.438(64)& 0.098(91)&   0.007(81)\\
 &      & 0.445(24)& 0.272(47)& 0.434(22)& 0.366(59)&$-$0.068(54)\\
\cline{2-7}
 &0.1600& 1.531(94)& 0.09 (11)& 1.512(86)& 0.22 (16)&$-$0.09 (14)\\
 &      & 0.407(27)& 0.276(61)& 0.395(26)& 0.44 (10)&$-$0.115(86)\\
\hline
(1.5\,,\,2)
 &0.1570& 1.167(38)& 0.209(81)& 1.185(37)& 0.086(80)&   0.119(87)\\
 &      & 0.597(25)& 0.472(64)& 0.587(22)& 0.548(60)&$-$0.075(78)\\
\cline{2-7}
 &0.1585& 1.213(50)& 0.19 (10)& 1.224(47)& 0.10 (12)&   0.08 (14)\\
 &      & 0.559(28)& 0.493(78)& 0.545(24)& 0.623(92)&$-$0.13 (12)\\
\cline{2-7}
 &0.1600& 1.283(67)& 0.17 (14)& 1.279(62)& 0.21 (20)&$-$0.04 (24)\\
 &      & 0.516(32)& 0.52 (10)& 0.496(29)& 0.77 (17)&$-$0.26 (20)\\
\hline
(0.9\,,\,2)
 &0.1570& 1.011(28)& 0.360(88)& 1.027(27)& 0.208(85)&   0.19 (13)\\
 &      & 0.685(28)& 0.753(90)& 0.690(26)& 0.713(75)&   0.05 (13)\\
\cline{2-7}
 &0.1585& 1.041(36)& 0.35 (11)& 1.056(35)& 0.19 (13)&   0.20 (21)\\
 &      & 0.647(33)& 0.79 (11)& 0.640(28)& 0.86 (12)&$-$0.09 (21)\\
\cline{2-7}
 &0.1600& 1.090(49)& 0.33 (15)& 1.096(48)& 0.26 (24)&   0.10 (35)\\
 &      & 0.599(37)& 0.85 (14)& 0.577(32)& 1.12 (23)&$-$0.36 (36)\\
\end{tabular}
\end{center}
\caption{ 
Parameters for the fit
$f^{-1}(q^2)=f^{-1}(q^2_{\rsub{max}})
 +c_1(q^2_{\rsub{max}}-q^2)+c_2(q^2_{\rsub{max}}-q^2)^2$,
where $c_2$ is set to zero for the linear fit.
For each $(m_Q,n)$ and $\kappa$, numbers in upper and lower rows 
correspond $f^0$ and $f^+$, respectively.
In all the cases, $\chi^2/$dof are less than unity. }
\label{tab:FF_fit}
\end{table}

\newlength{\figwidth}
\setlength{\figwidth}{0.72\textwidth}
\addtolength{\figwidth}{-0.5\columnsep}

\begin{figure}[p]
\begin{center}
\vspace*{-0.8cm}
\leavevmode\psfig{file=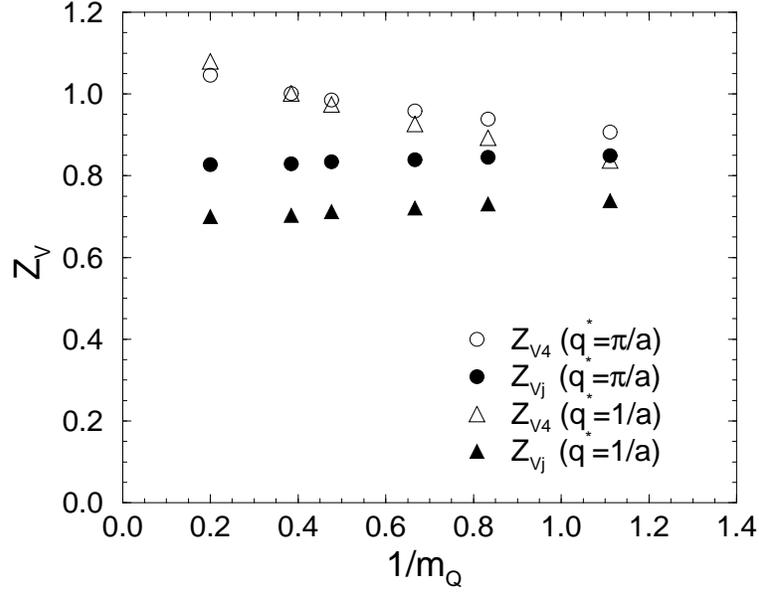,width=\figwidth}
\vspace{-1cm}
\end{center}
\caption{ 
Renormalization constant for the vector current with two scales for
the coupling constant, $q^{\star}=\pi/a$ and $1/a$.  
The open and filled symbols represent $Z_{V_4}$ and
$Z_{V_j}$, respectively. }
\label{fig:Z_V}
\end{figure}

\begin{figure}[p]
\begin{center}
\vspace*{0.8cm}
$\kappa=0.1570$  \vspace{-0.8cm} \\
\leavevmode\psfig{file=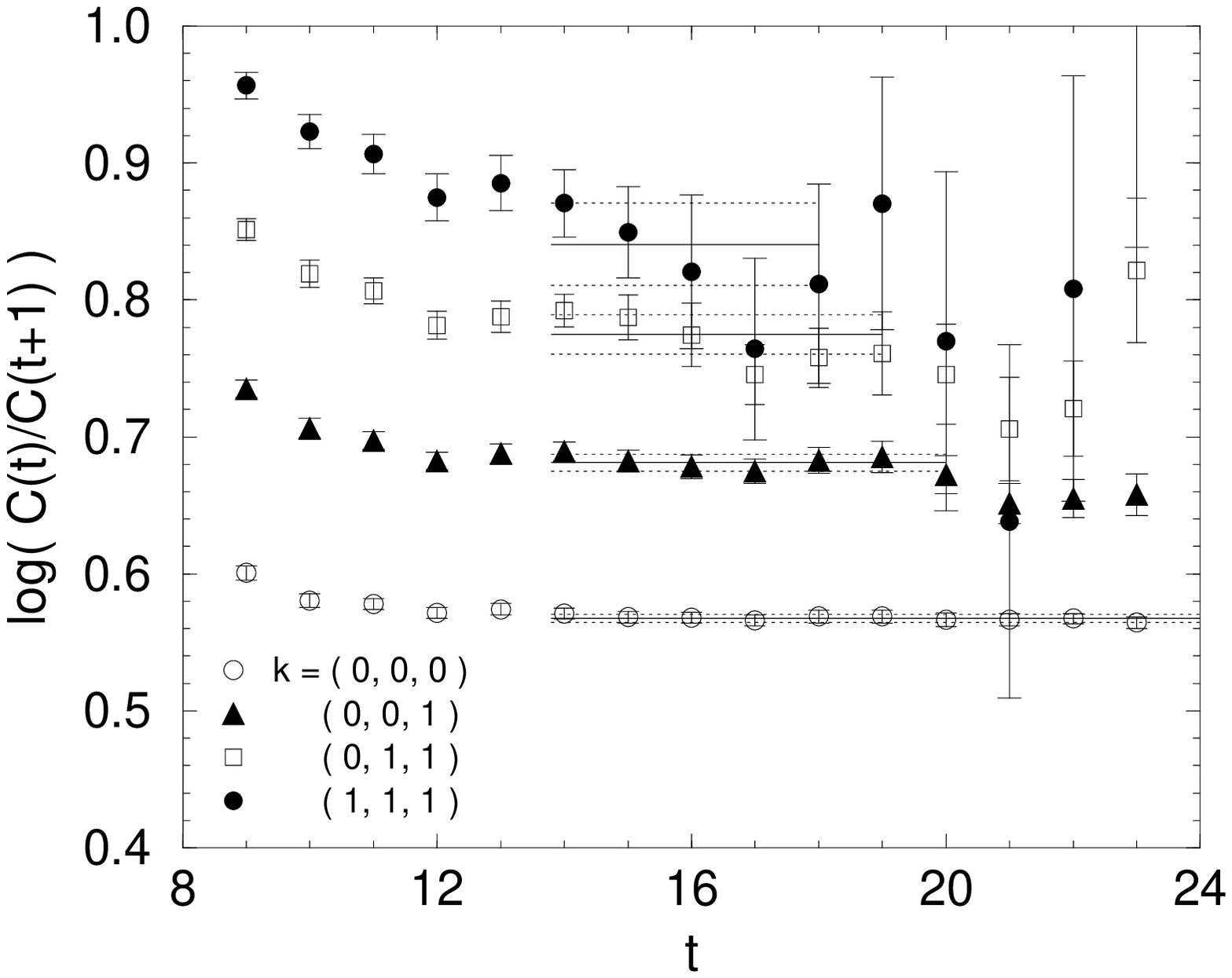,width=\figwidth}
\vspace{1.8cm}\\
$\kappa=0.1600$  \vspace{-0.8cm} \\
\leavevmode\psfig{file=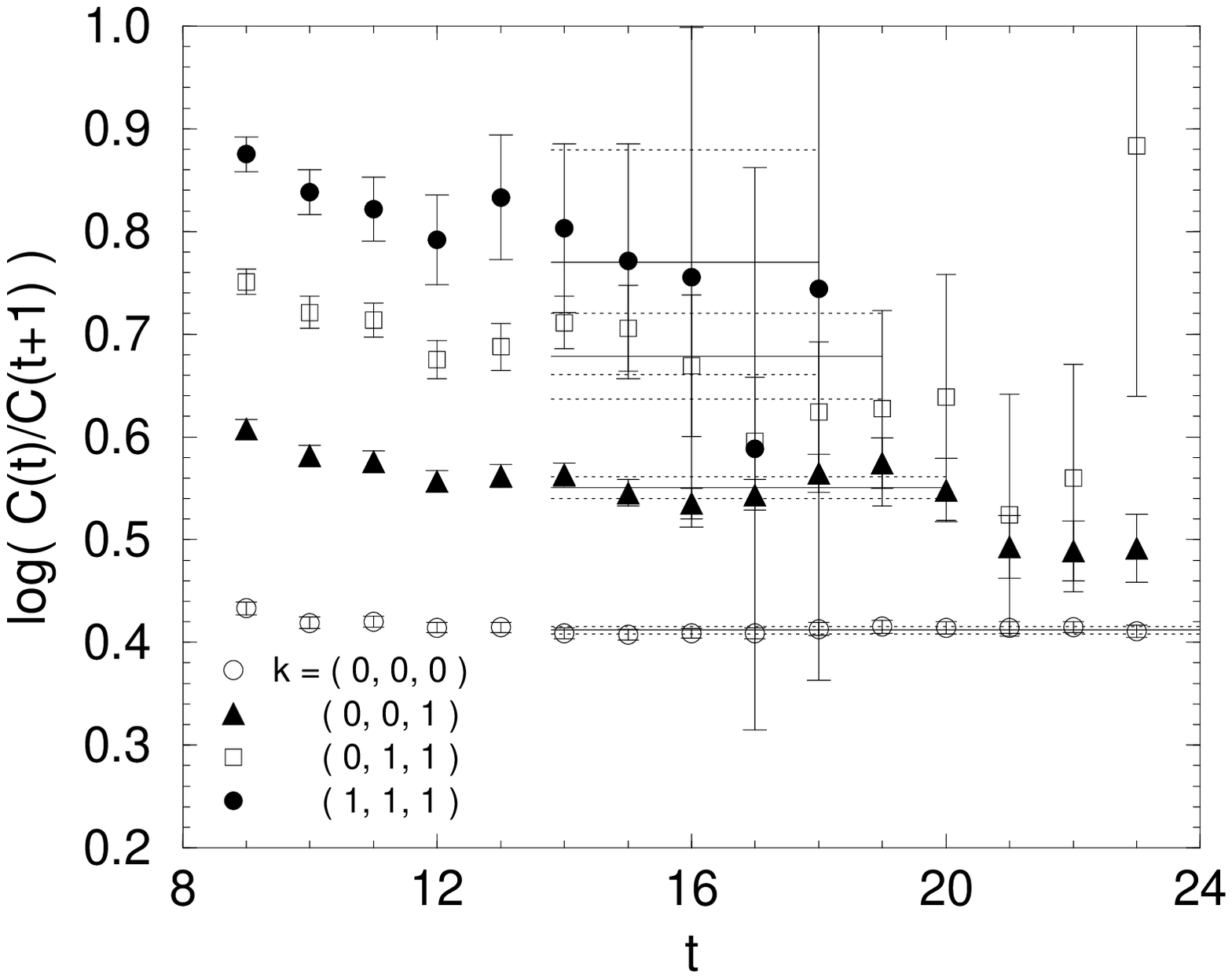,width=\figwidth}
\vspace{-1cm}
\end{center}
\caption{
Effective mass plot of pion at $\kappa=0.1570$ and $0.1600$.
The horizontal solid lines represent the fitted values and the fitting
range with the statistical errors (dotted lines).}
\label{fig:EP_pi}
\end{figure}

\begin{figure}[p]
\begin{center}
\vspace*{-0.8cm}
\leavevmode\psfig{file=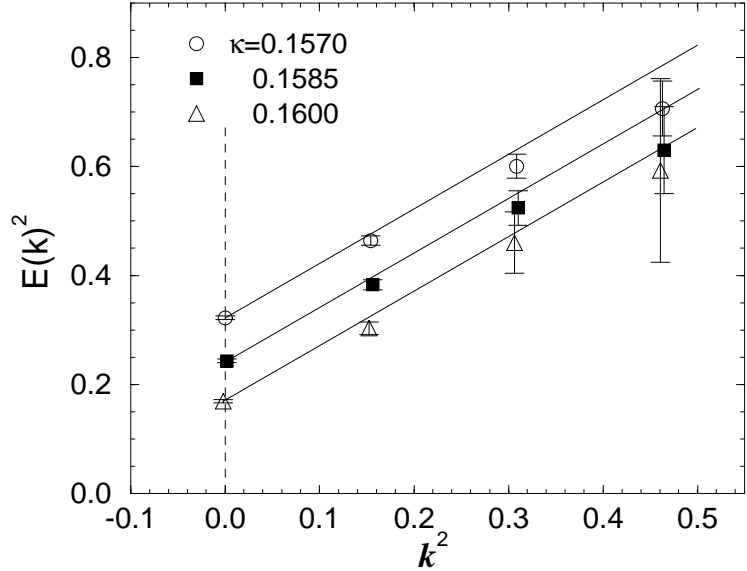,width=\figwidth}
\vspace{-1cm}
\end{center}
\caption{ Dispersion relation for pion. 
The solid lines represent the relation 
$E_{\pi}^{\,2}(\vec{k})=m_{\pi}^{\,2} + \vec{k}^2$ 
with $m_{\pi}$ the rest mass obtained in the simulation.
For $\kappa=0.1585$ and $0.1600$, symbols are slightly shifted
in horizontal direction for clarity. }
\label{fig:DR_pi}
\end{figure}

\begin{figure}[p]
\begin{center}
\vspace*{-0.8cm}
$m_Q=2.6$, $\kappa=0.1570$ \vspace{-0.8cm} \\
\leavevmode\psfig{file=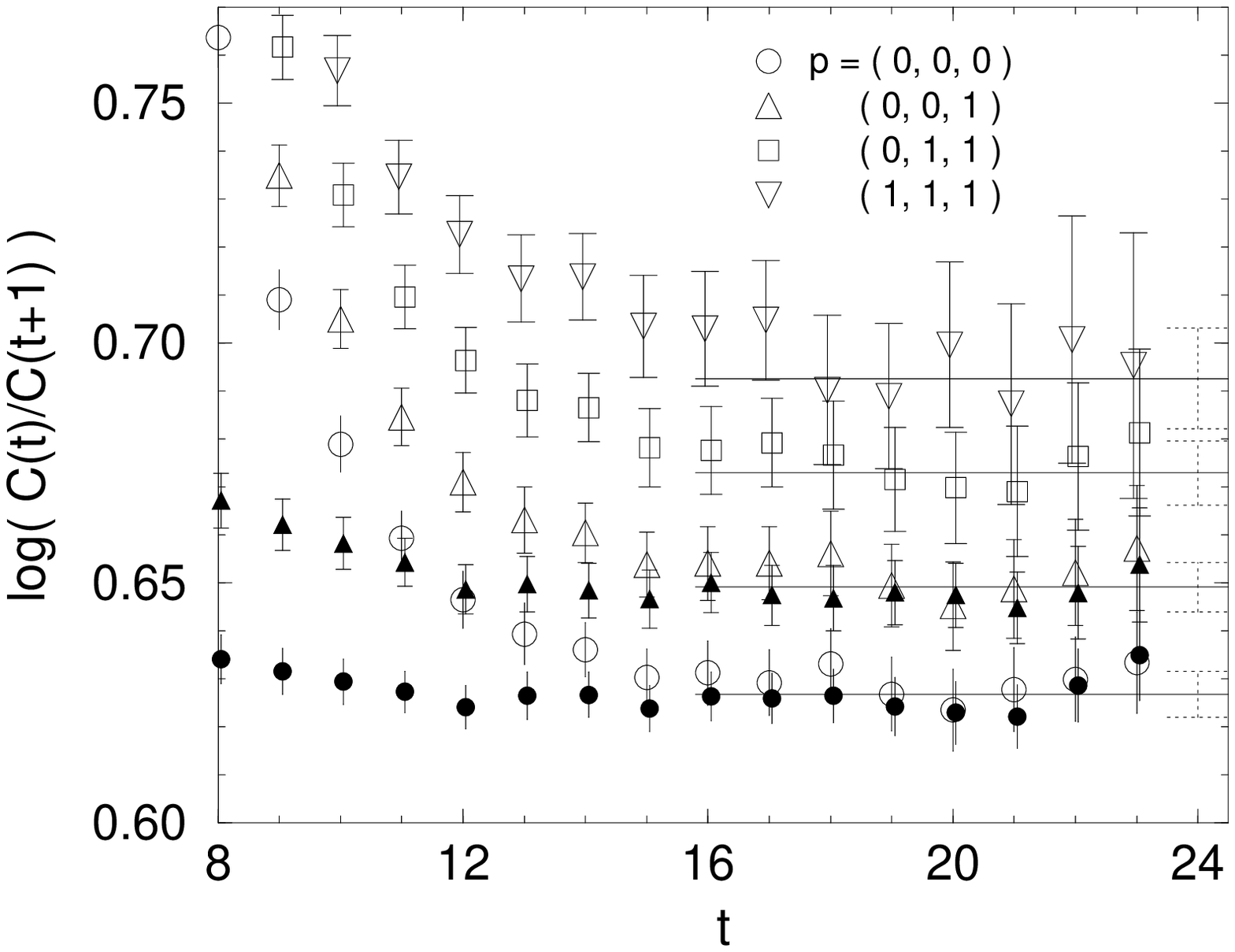,width=\figwidth}
\vspace{-0.8cm} \\
$m_Q=2.6$, $\kappa=0.1600$ \vspace{-0.8cm} \\
\leavevmode\psfig{file=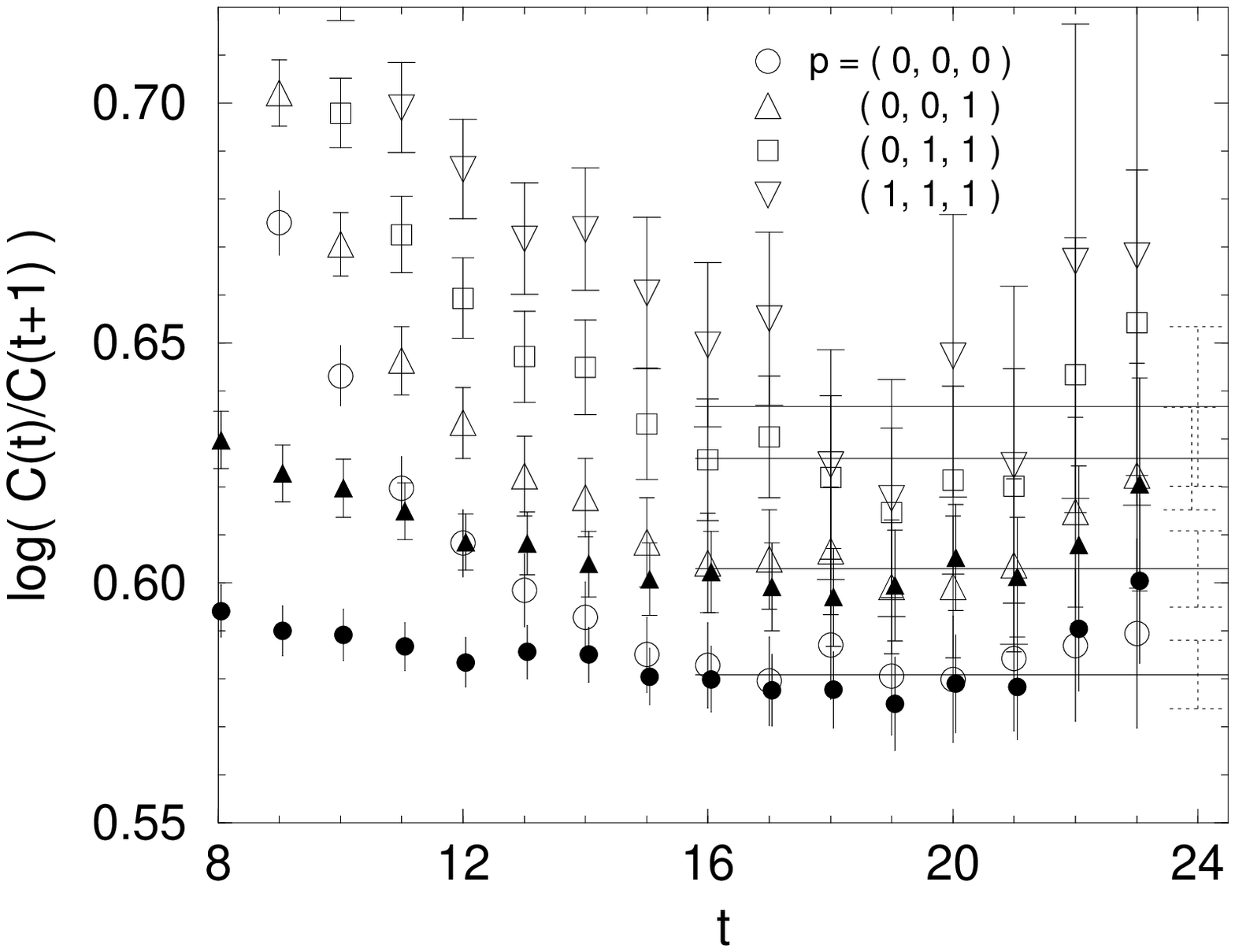,width=\figwidth}
\vspace{-1cm}
\end{center}
\caption{
Effective mass plot of $B$ meson at $m_Q=2.6$ and
$\kappa=0.1570$, $0.1600$.
Results with the smeared source (filled symbols) are shown for 
$|\vec{p}|^2=0, 1$ as well as the results with the local source (open
symbols).
The horizontal solid lines express the average values over the
results of single exponential fit of the local-local and 
the smeared-local correlation functions.
The statistical errors of the fitted values are displayed
at the right end of the lines.
For all $m_Q$, $\kappa$, and momentum, the fit ranges are set to
$t=16-24$.}

\label{fig:EP_B}
\end{figure}

\begin{figure}[p]
\begin{center}
\vspace*{-0.8cm}
\leavevmode\psfig{file=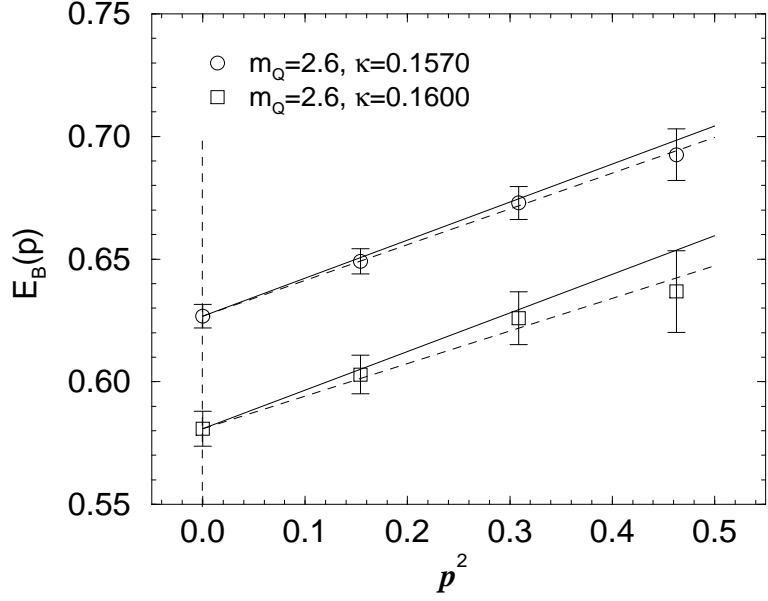,width=\figwidth}
\vspace{-1cm}
\end{center}
\caption{ Dispersion relation for the $B$ meson 
at $m_Q=2.6$ and $\kappa=0.1570$, $0.1600$. 
The solid lines represent the relation 
$E_{\bar{q}Q}(\vec{p})=E_{\bar{q}Q}(0)+\vec{p}^2/2m_B$,
for which $m_B$ is determined with the tree level formula
$m_B=m_Q+E_{\bar{q}Q}(0)$. 
Dashed lines represent the same relation with the renormalized
$m_B$ at the scale $q^{\star}=1/a$. }
\label{fig:DR_B}
\end{figure}

\clearpage 

\begin{figure}[p]
\begin{center}
$m_Q=2.6$, $\kappa=0.1570$, $C_4^{(3)}$ \vspace{-0.8cm} \\
\leavevmode\psfig{file=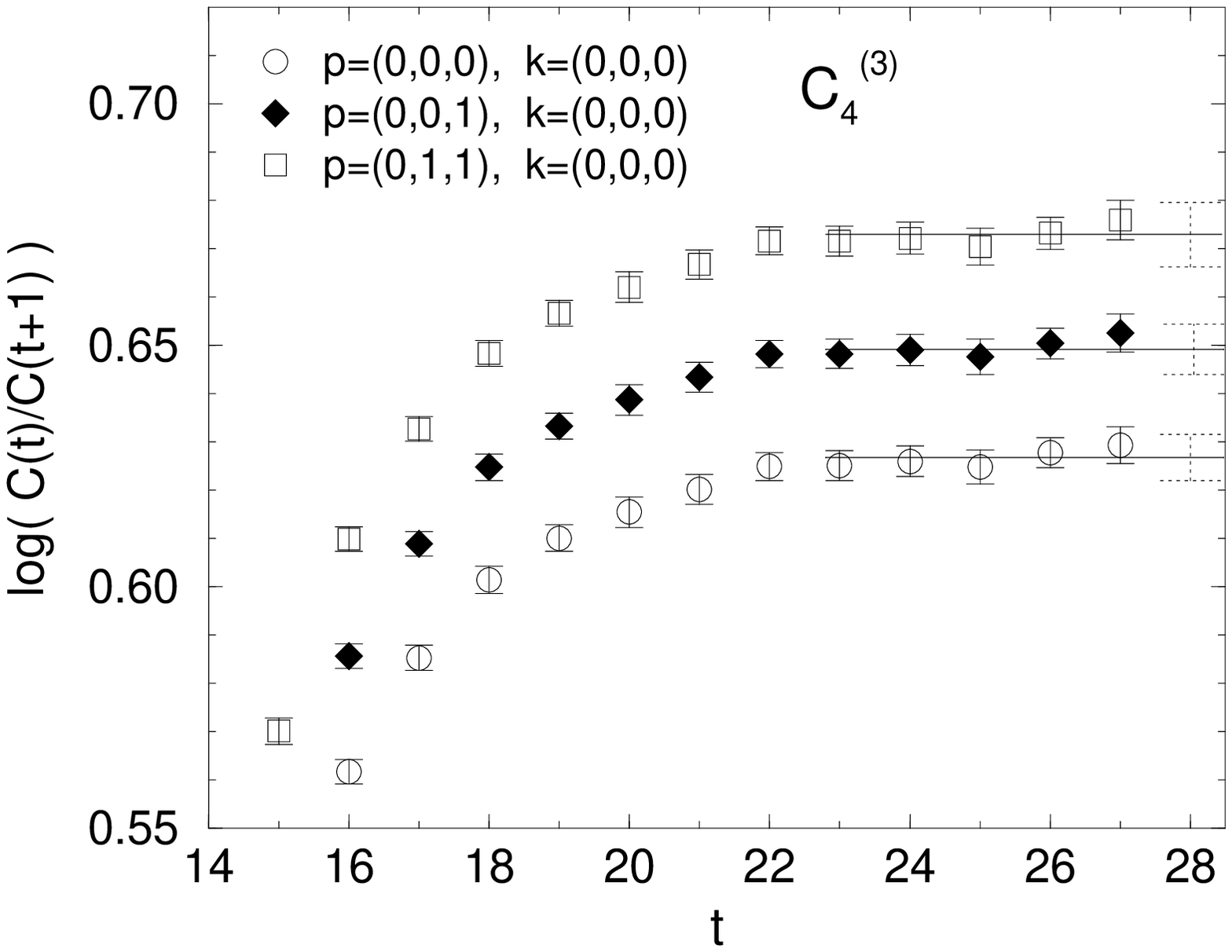,width=\figwidth}
\vspace{-0.8cm}\\
$m_Q=2.6$, $\kappa=0.1570$, $C_4^{(3)}$ \vspace{-0.8cm} \\
\leavevmode\psfig{file=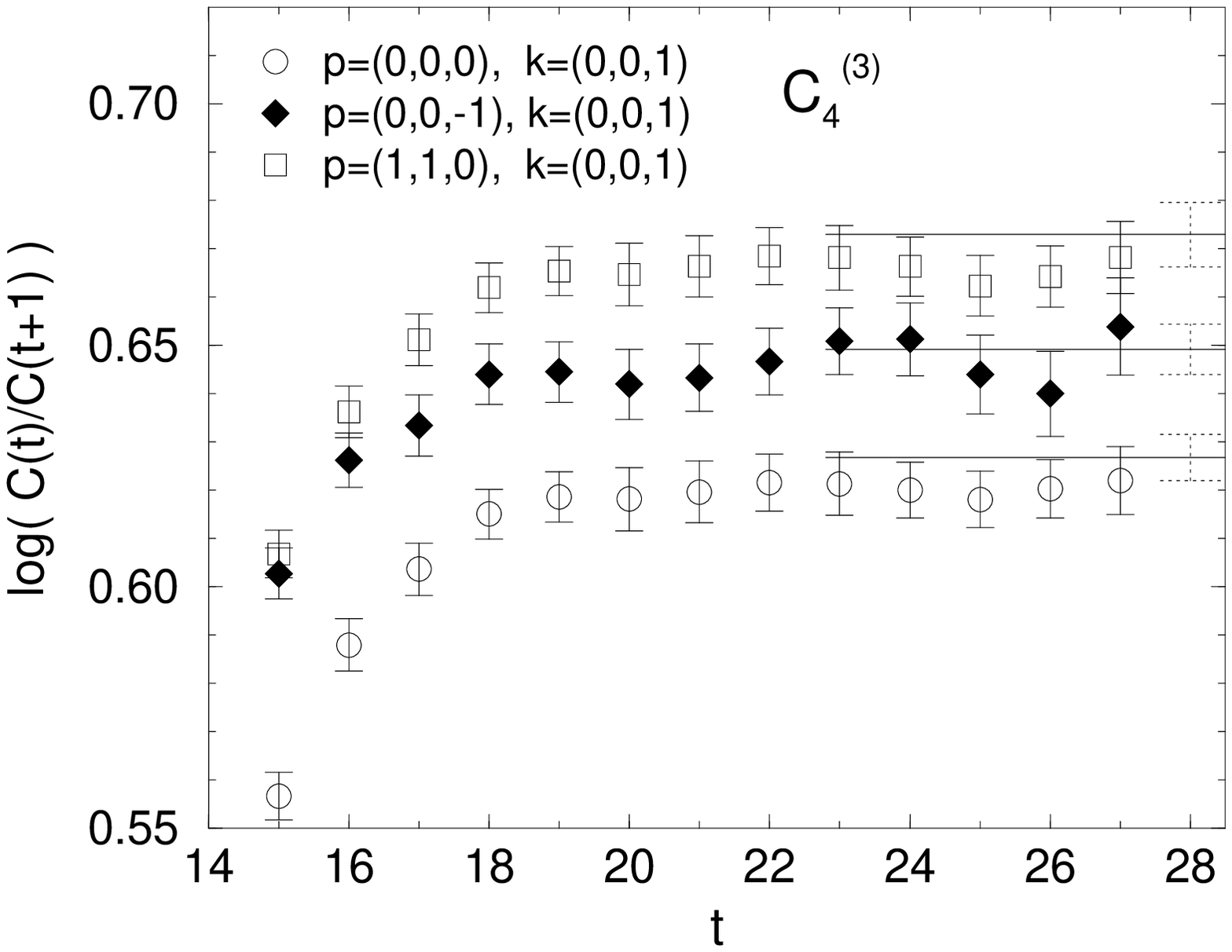,width=\figwidth}
\end{center}
\clearpage
\begin{center}
$m_Q=2.6$, $\kappa=0.1570$, \ $\vec{k}\!\cdot\!\vec{C}^{(3)}$
\vspace{-0.8cm} \\
\leavevmode\psfig{file=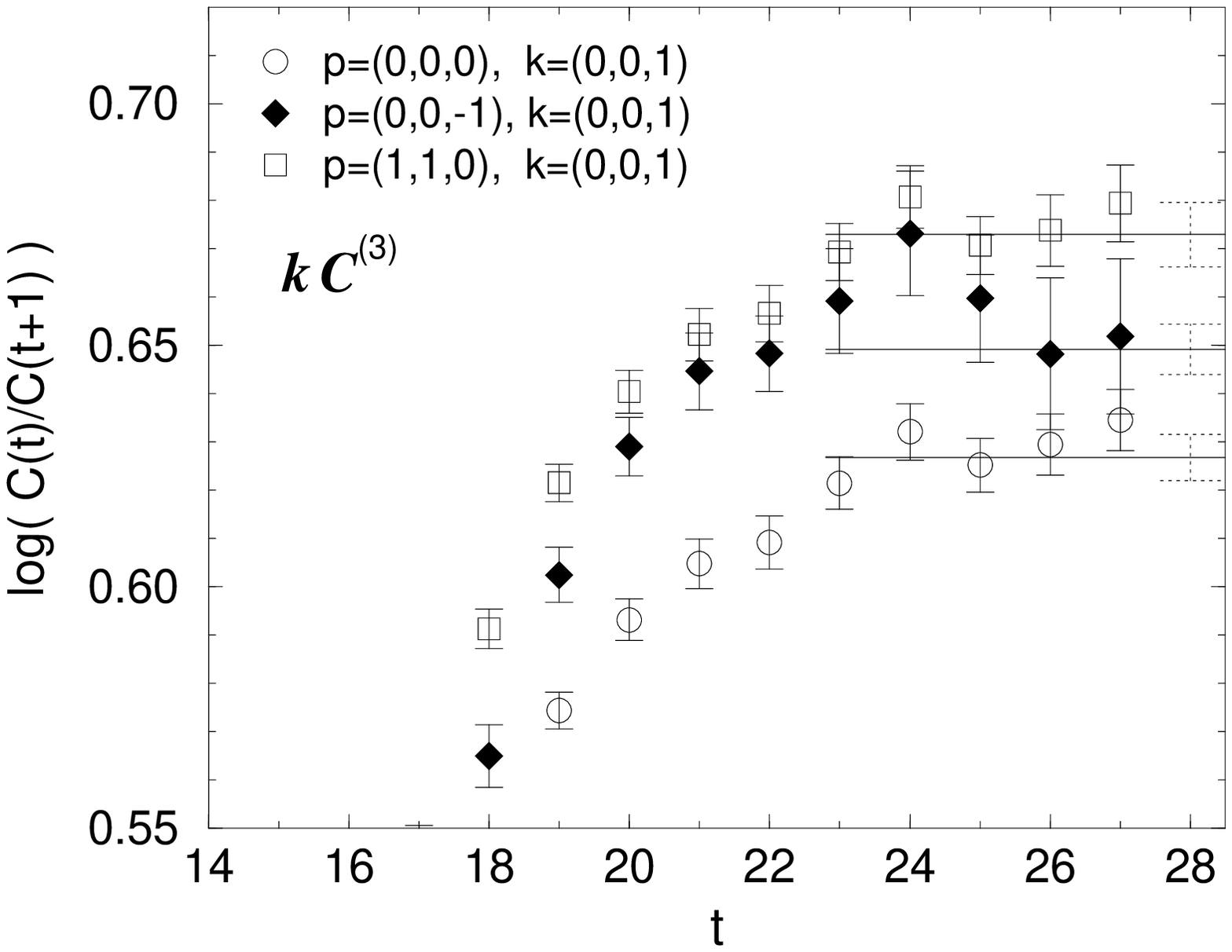,width=\figwidth}
\vspace{-0.8cm}\\
$m_Q=2.6$, $\kappa=0.1600$, $C_4^{(3)}$ \vspace{-0.8cm} \\
\leavevmode\psfig{file=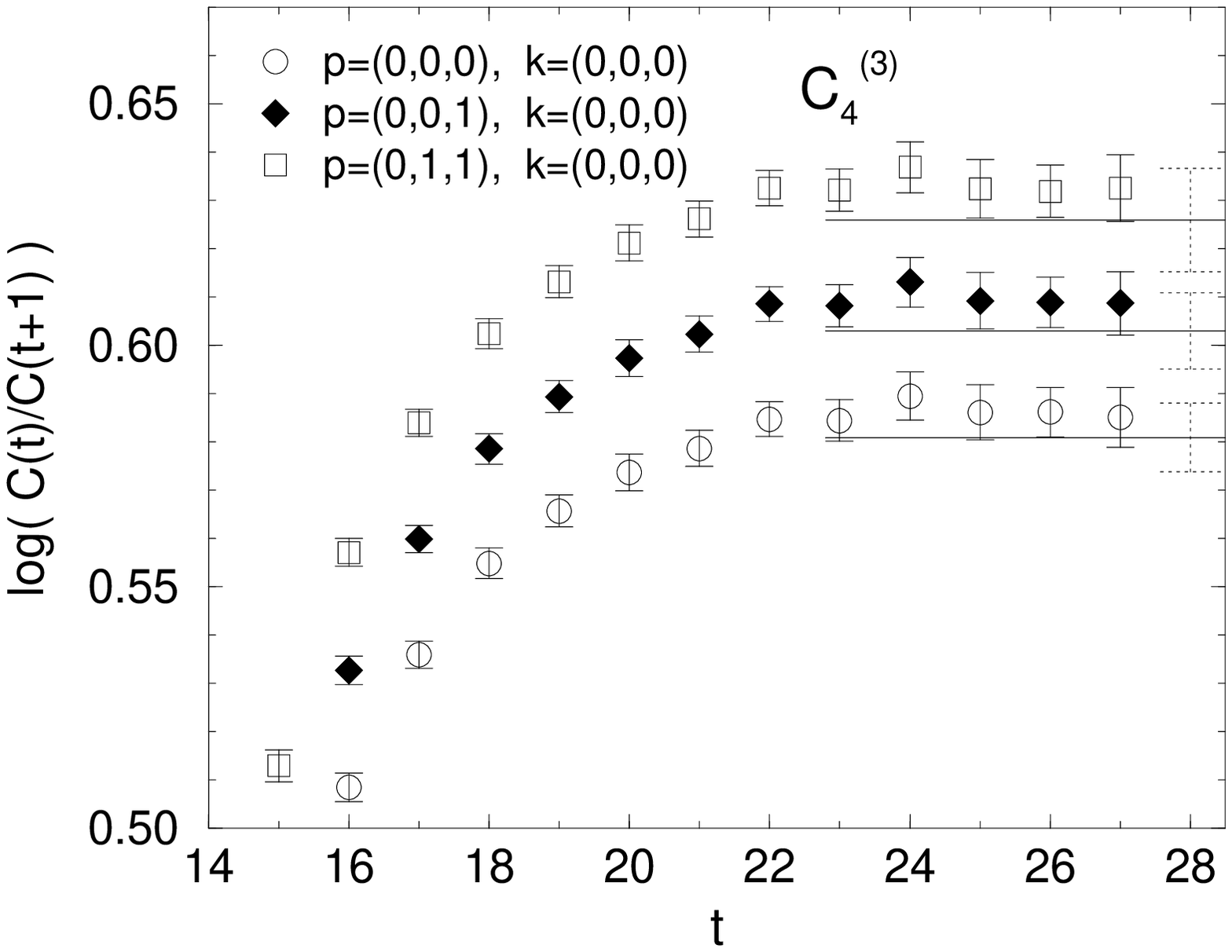,width=\figwidth}
\vspace{-1cm}
\end{center}
\caption{
Effective mass plot for the three-point functions at $m_Q=2.6$ 
and $\kappa=0.1570$, $0.1600$.
The horizontal lines express the values obtained from the two-point
correlation functions with the statistical errors indicated
at the right end of the lines.
Top two figures are for $C_4^{(3)}$ and the third is for
$\vec{k}\!\cdot\!\vec{C}^{(3)}$ at $\kappa=0.1570$, and bottom
figure is for $C_4^{(3)}$ at $\kappa=0.1600$.}
\label{fig:EP_C3}
\end{figure}

\begin{figure}[p]
\begin{center}
\vspace*{-0.8cm}
\leavevmode\psfig{file=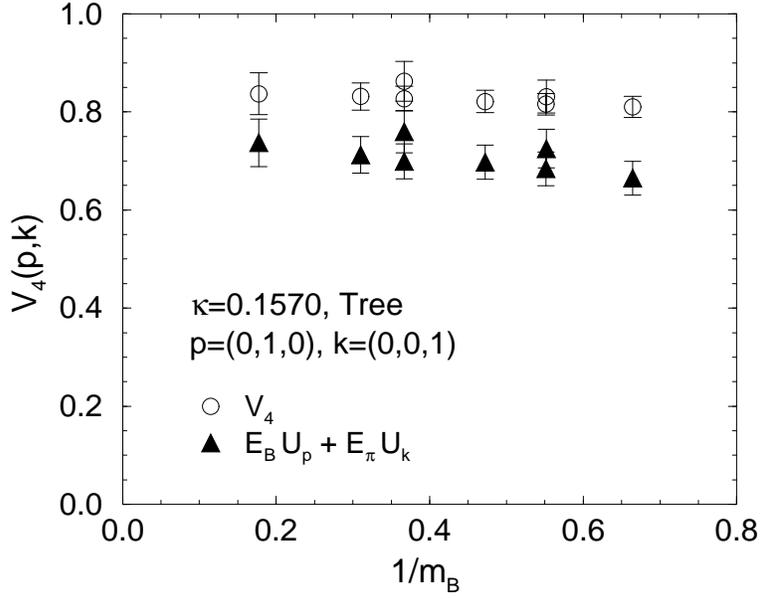,width=\figwidth}
\vspace{-1cm}
\end{center}
\caption{ Comparison of $\hat{V}_4$ to
$(\, E_B \hat{U}_p + E_{\pi} \hat{U}_k\,)$ 
for $i_q=6$ at $\kappa=0.1570$.  }
\label{fig:V4condition}
\end{figure}

\begin{figure}[p]
\begin{center}
\vspace*{-0.8cm}
\leavevmode\psfig{file=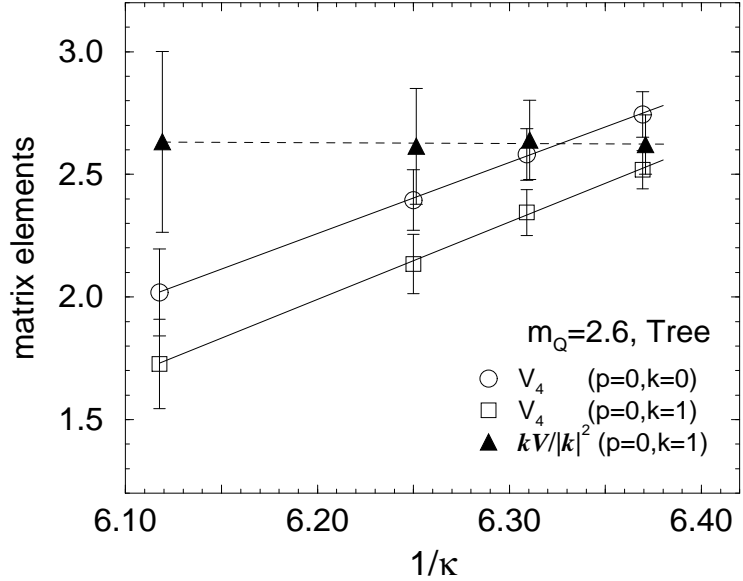,width=\figwidth}
\vspace{-1cm}
\end{center}
\caption{
Chiral extrapolation of the matrix elements for $m_Q=2.6$.
$V_4$ and $\vec{k}\cdot\vec{V}/|\vec{k}|^2$ are shown
for two momentum configurations $i_q=1$, $2$. 
The solid and the dashed lines represent the linear fit.
}
\label{fig:Chiral_ext}
\end{figure}

\begin{figure}[p]
\begin{center}
\vspace*{-0.8cm}
\leavevmode\psfig{file=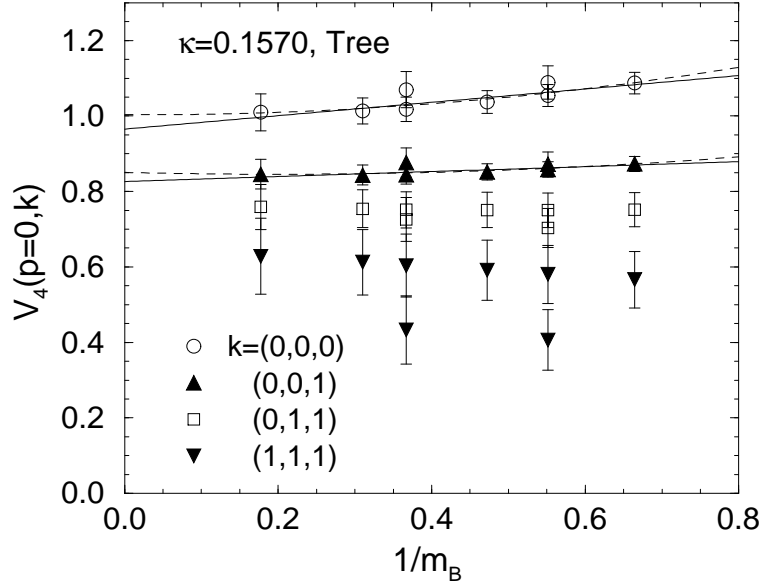,width=\figwidth}
\vspace{-1cm}
\end{center}
\caption{ $\hat{V}_4$ at $\kappa=0.1570$.
For $\vec{k}=0$ and $|\vec{k}|=1$, the solid and the dashed 
lines represent the results of linear and quadratic fits,
respectively. }
\label{fig:V4_kdep}
\end{figure}

\begin{figure}[p]
\begin{center}
\vspace*{-0.8cm}
\leavevmode\psfig{file=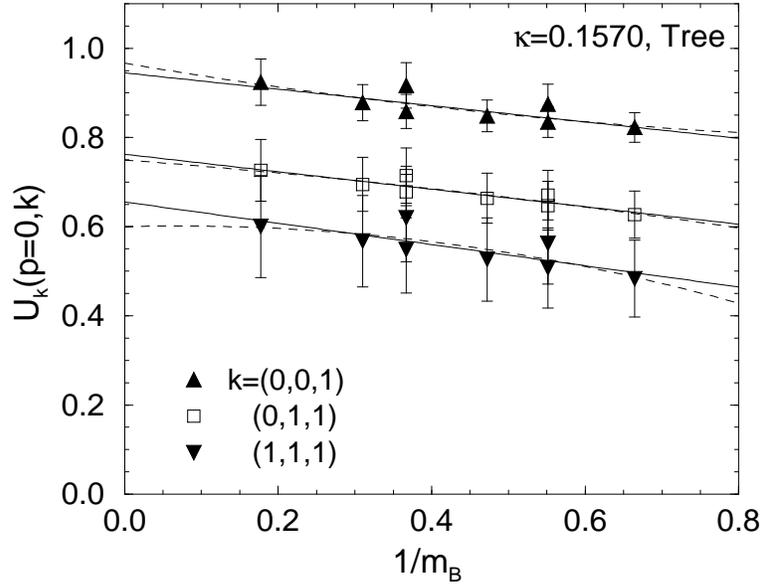,width=\figwidth}
\vspace{-1cm}
\end{center}
\caption{ $\hat{U}_k$ at $\kappa=0.1570$.
The solid and the dashed lines represent the results of 
linear and quadratic fits, respectively. }
\label{fig:Vk_kdep}
\end{figure}

\begin{figure}[p]
\begin{center}
\vspace*{-0.8cm}
\leavevmode\psfig{file=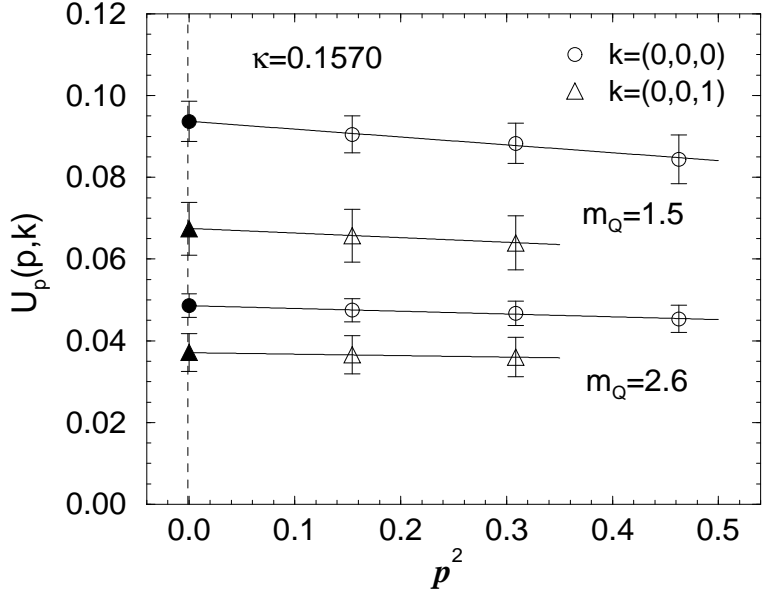,width=\figwidth}
\vspace{-1cm}
\end{center}
\caption{ Extraction of $\hat{U}_p(\vec{p}\!=\!0,\vec{k})$ is shown 
for $m_Q=2.6$ and $1.5$ at $\kappa=0.1570$.
The extrapolation is carried out linearly in $\vec{p}^2$.
For $\vec{k}=0$, $i_q=5, 13, 20$ are used.
For $\vec{k}=1$,  $\hat{U}_p(0,\vec{k})$ is determined 
using $i_q=6, 14$, for which $\vec{p}$ and $\vec{k}$ are perpendicular.  }
\label{fig:Vp_ext}
\end{figure}

\begin{figure}[p]
\vspace*{-0.8cm}
\begin{center}
\leavevmode\psfig{file=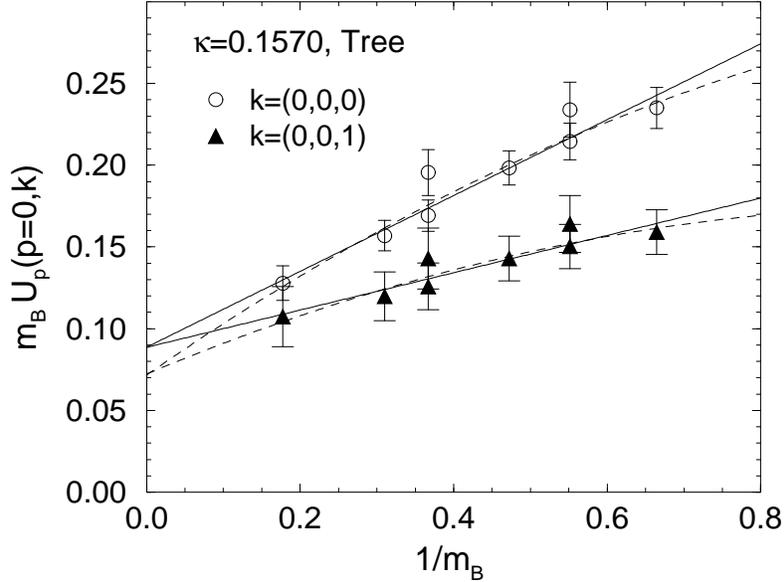,width=\figwidth}
\vspace{-1cm}
\end{center}
\caption{ $\hat{U}_p$ multiplied by $m_B$ at $\kappa=0.1570$ as a
function of $1/m_B$.
The values of $m_B$ are determined with the tree level formula.
The solid and the dashed lines represent the linear and
the quadratic fits, respectively. } 
\label{fig:Vp_kdep}
\end{figure}

\begin{figure}[p]
\begin{center}
\vspace*{-0.8cm}
\leavevmode\psfig{file=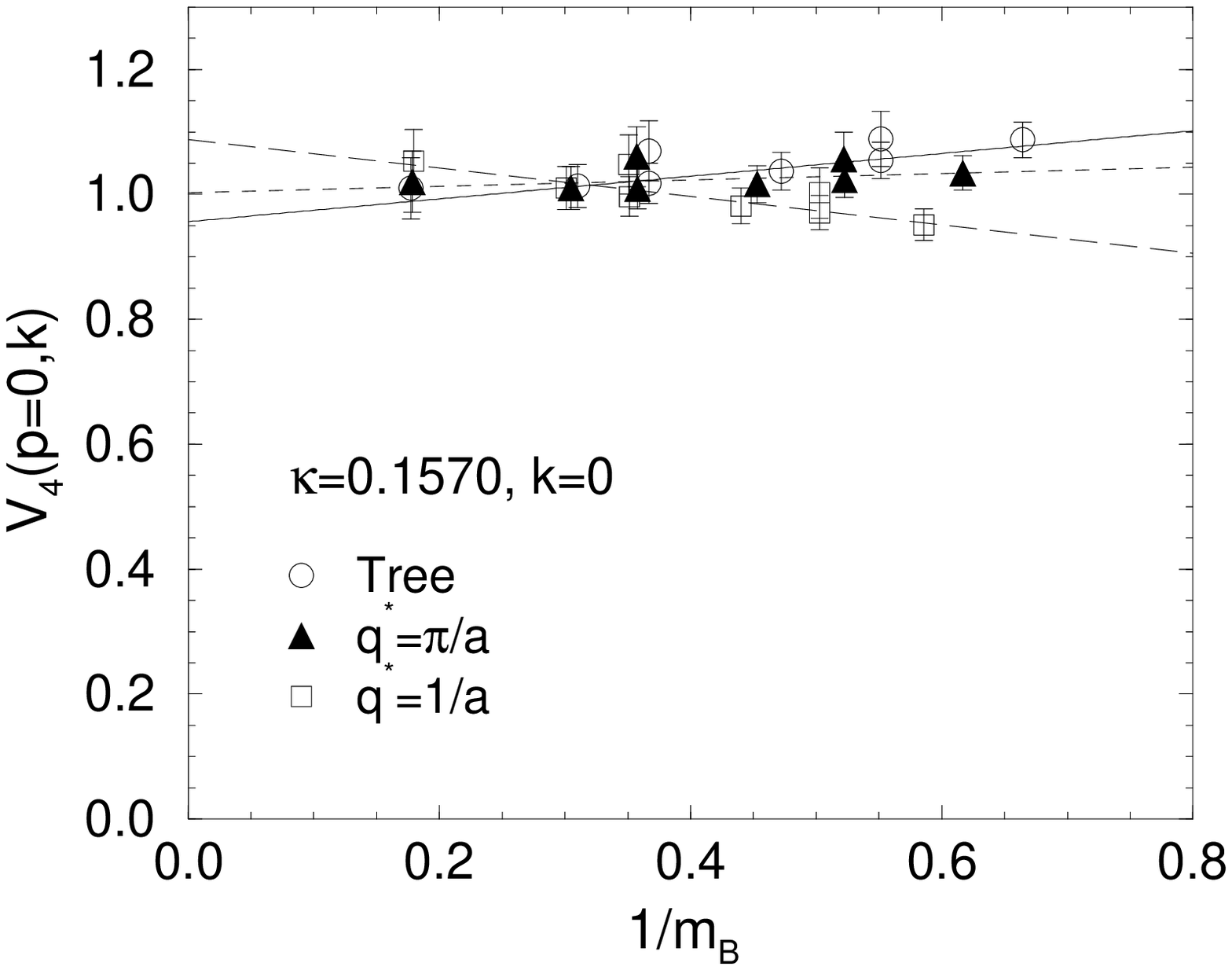,width=\figwidth}
\vspace{-1.8cm}\\
\leavevmode\psfig{file=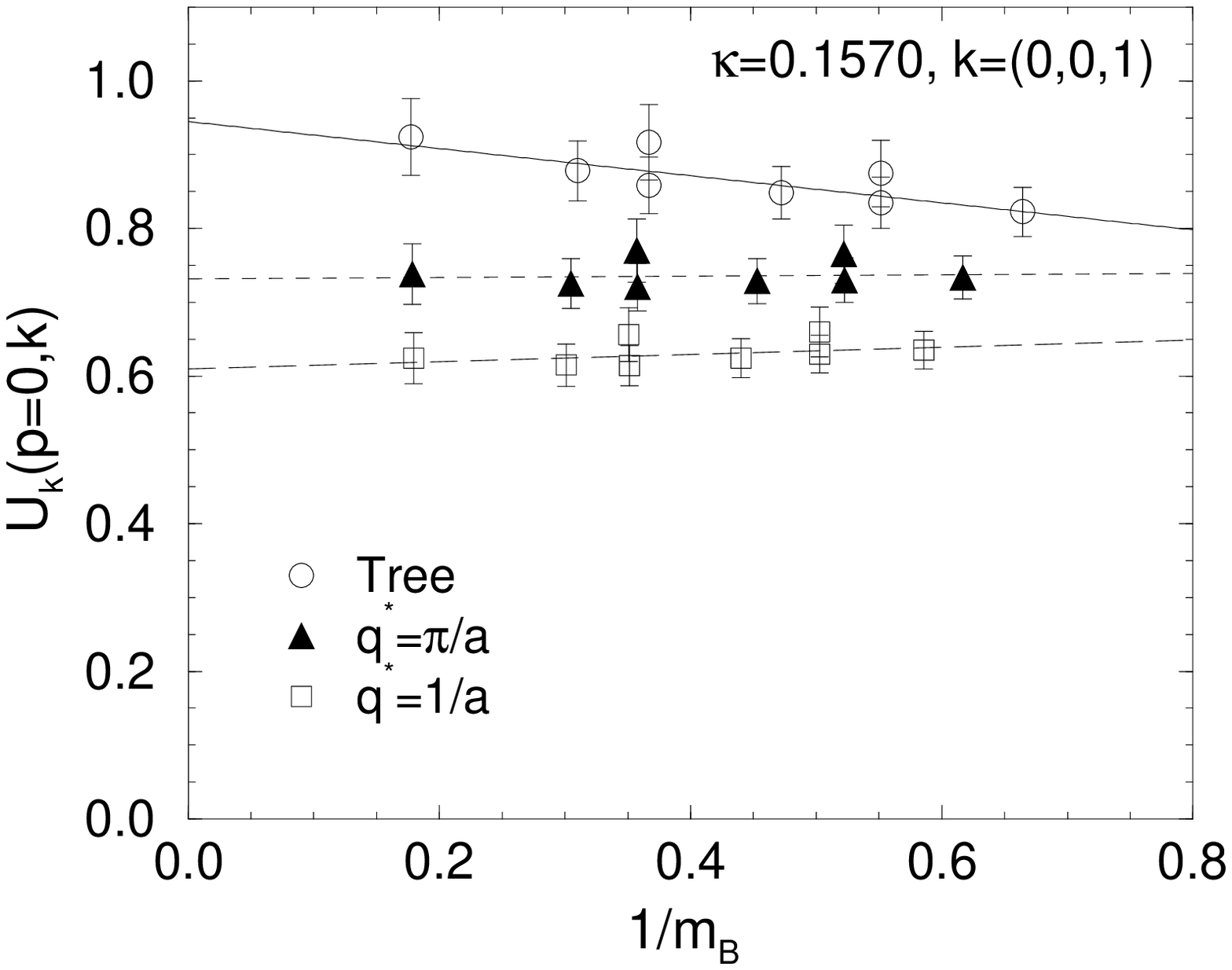,width=\figwidth}
\vspace{-1.8cm}\\
\leavevmode\psfig{file=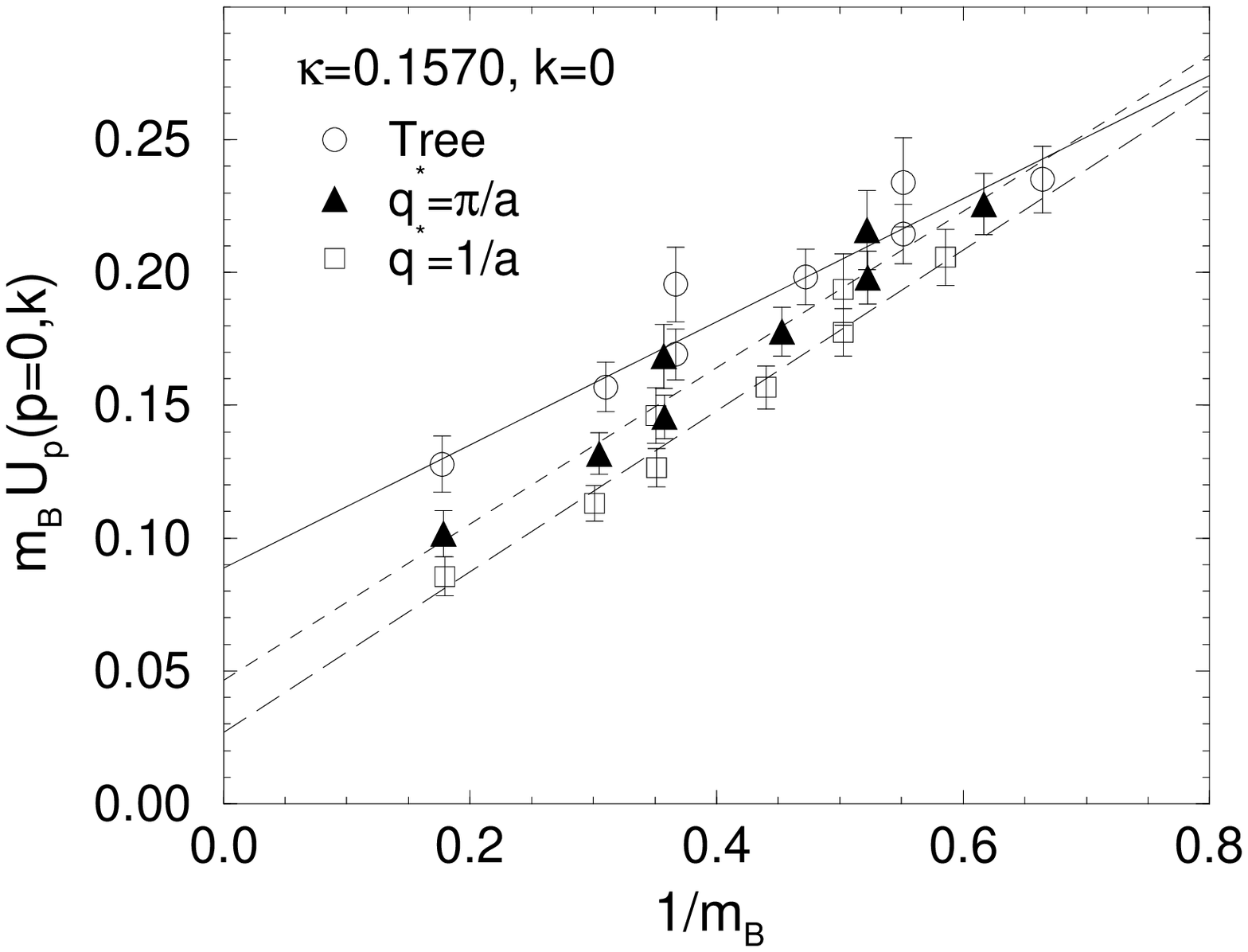,width=\figwidth}
\vspace{-1cm}
\end{center}
\caption{ 
One-loop renormalized $\hat{V}_4$, $\hat{U}_k$, and $\hat{U}_p$ as a
function of $1/m_B$.
The solid, the dashed, and the long dashed lines represent
the linear fits. }
\label{fig:V_renorm}
\end{figure}

\clearpage

\begin{figure}[p]
\begin{center}
\vspace*{-0.8cm}
\leavevmode\psfig{file=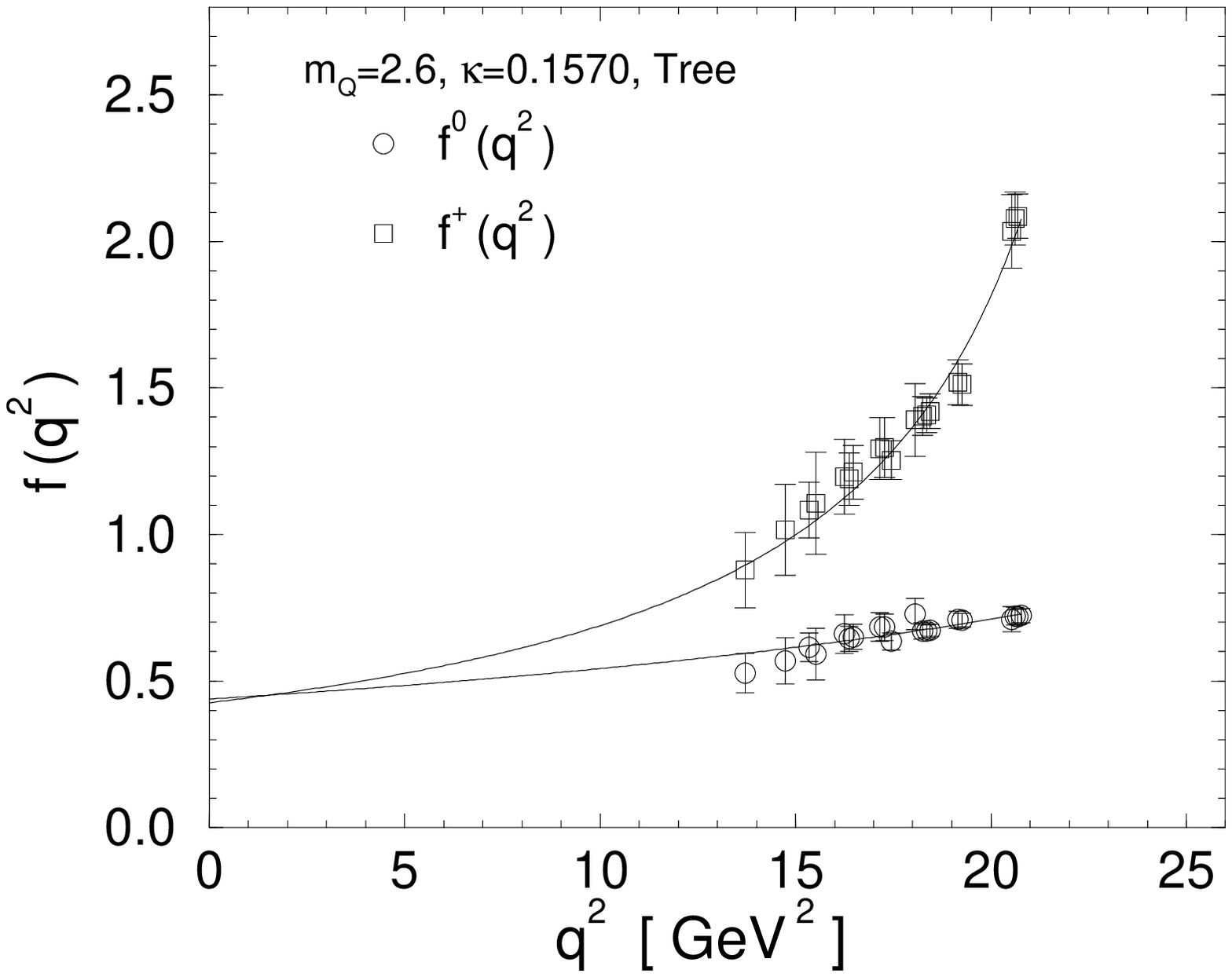,width=\figwidth}
\vspace{-1.8cm}\\
\leavevmode\psfig{file=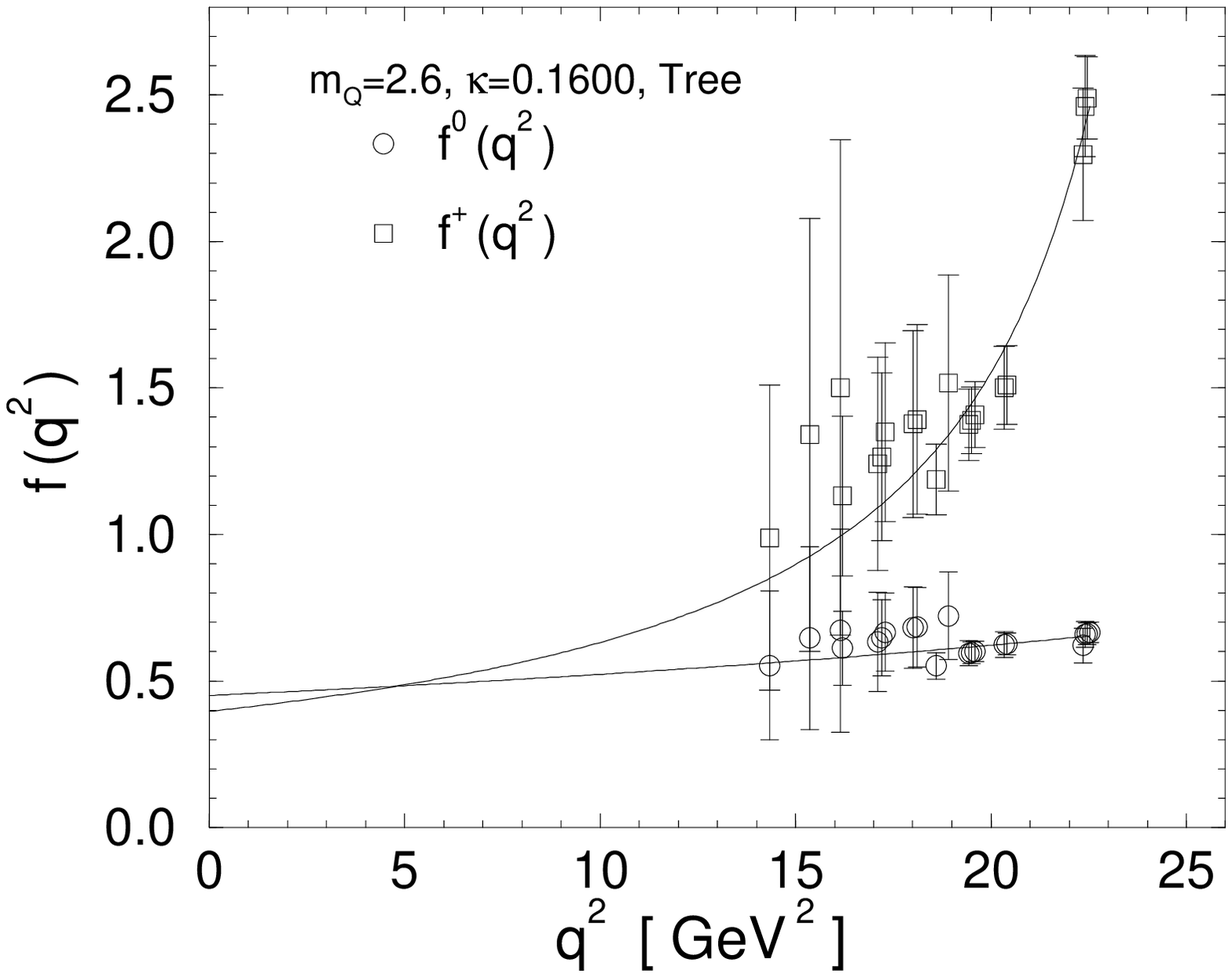,width=\figwidth}
\vspace{-1cm}
\end{center}
\caption{ 
Form factors at $m_Q=2.6$ and $\kappa=0.1570$, $0.1600$. 
The solid curves represent the fit to single pole functions. }
\label{fig:FF_02}
\end{figure}

\begin{figure}[p]
\begin{center}
\vspace*{-0.8cm}
\leavevmode\psfig{file=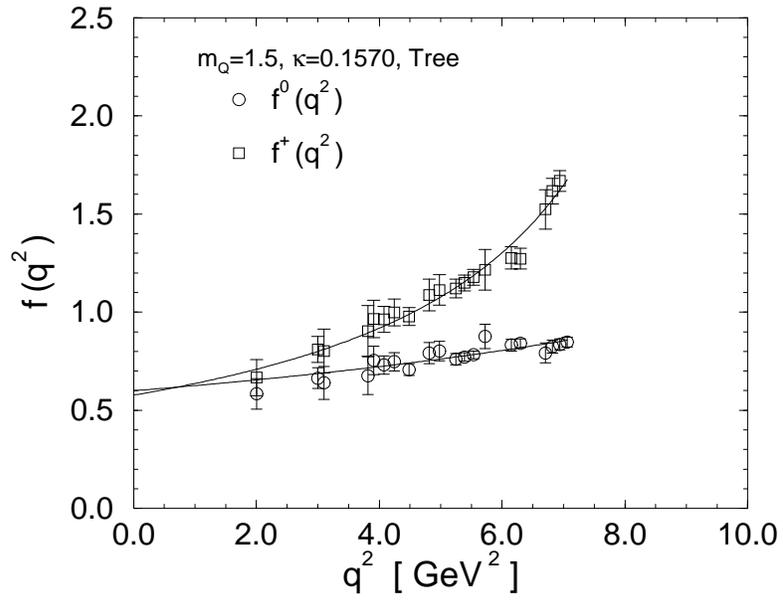,width=\figwidth}
\vspace{-1cm}\\
\leavevmode\psfig{file=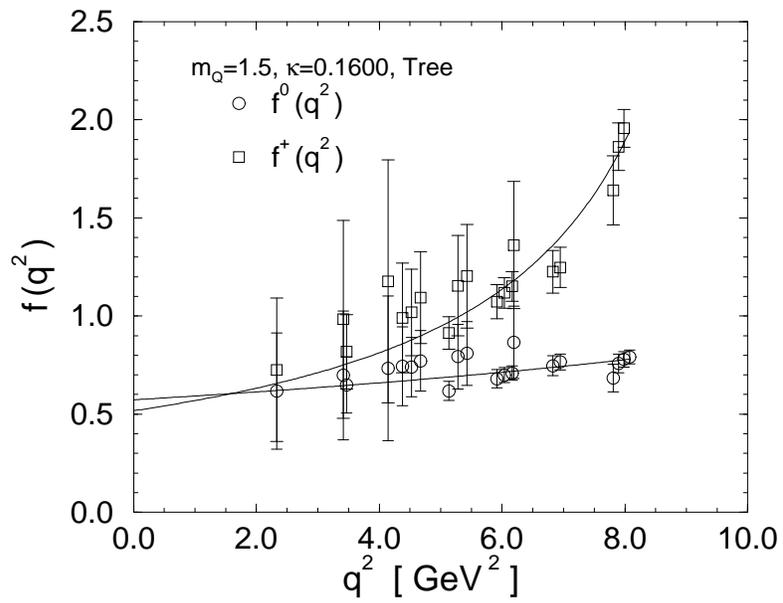,width=\figwidth}
\vspace{-1cm}
\end{center}
\caption{ 
Form factors at $m_Q=1.5$ and $\kappa=0.1570$, $0.1600$.  }
\label{fig:FF_04}
\end{figure}

\begin{figure}[p]
\begin{center}
\vspace*{-0.8cm}
\leavevmode\psfig{file=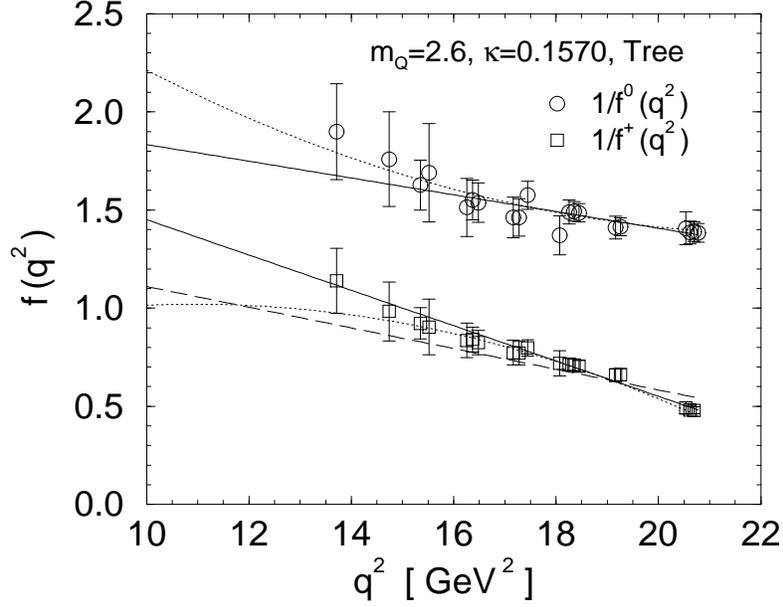,width=\figwidth}
\vspace{-1cm}
\end{center}
\caption{ 
$1/f^0$ and $1/f^+$ as a function of $q^2$ at $m_Q=2.6$ and
$\kappa=0.1570$. 
The solid and the dotted curves represent the linear and the
quadratic fits, respectively.
The long dashed line represents the linear fit with the constraint
$m_{\rsub{pole}}=m_{B^*}$, where $m_B^*$ is the $B^*$ meson
mass obtained from the two-point correlation function.
} 
\label{fig:FFr_02}
\end{figure}

\begin{figure}[p]
\begin{center}
\vspace*{-0.8cm}
\leavevmode\psfig{file=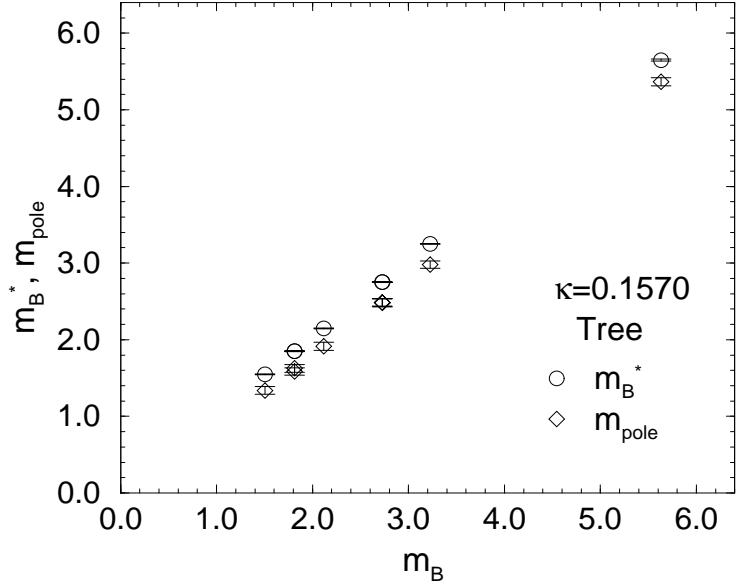,width=\figwidth}
\vspace{-1cm}
\end{center}
\caption{ 
$m_{B^*}$ obtained from the two-point correlation function of the
$B^*$ meson and the pole mass from the linear fit of $1/f^+$. } 
\label{fig:pole_mass}
\end{figure}

\begin{figure}[p]
\begin{center}
\vspace*{-0.8cm}
\leavevmode\psfig{file=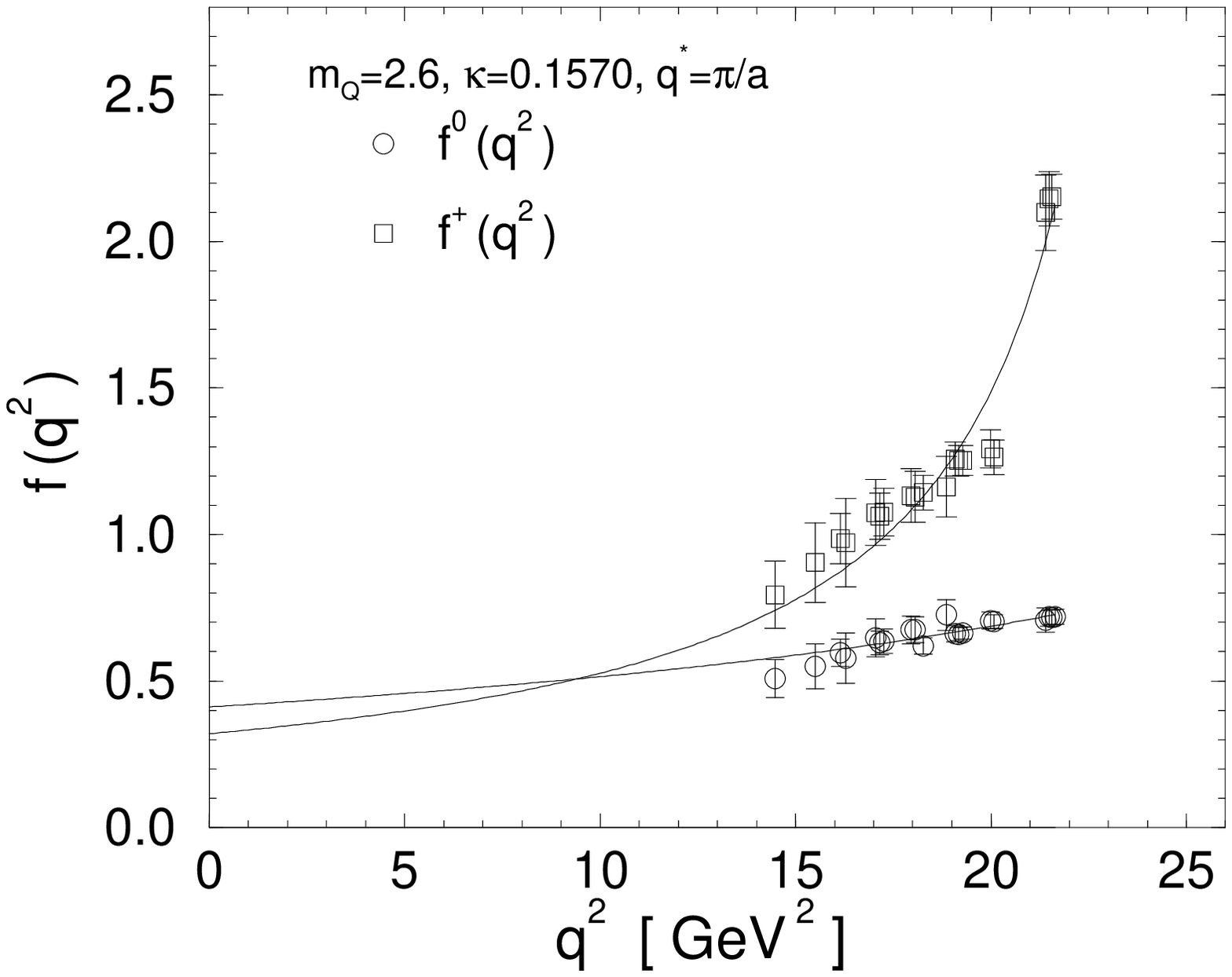,width=\figwidth}
\vspace{-1.8cm}\\
\leavevmode\psfig{file=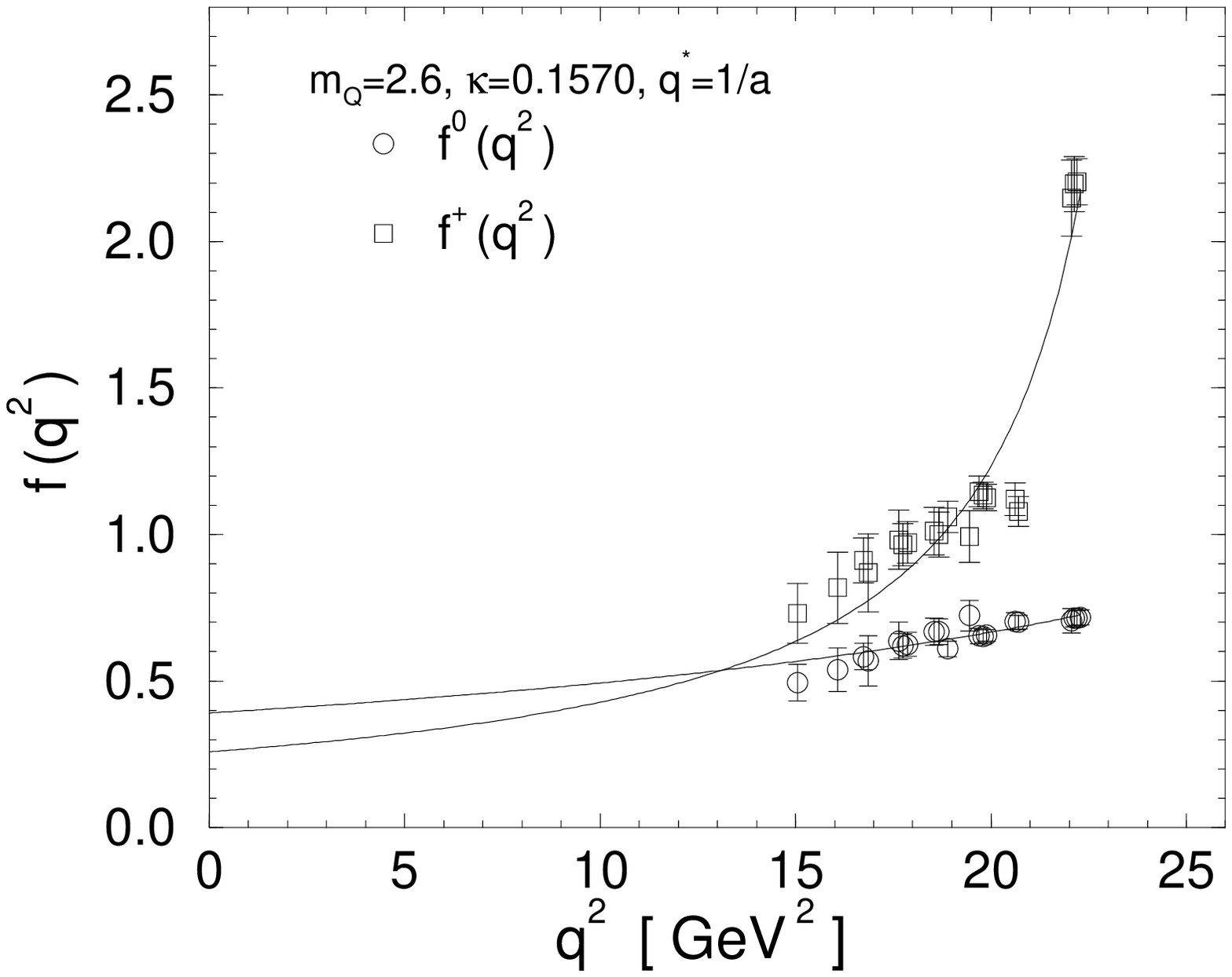,width=\figwidth}
\vspace{-1cm}
\end{center}
\caption{ 
Renormalized form factors at $m_Q=2.6$ and $\kappa=0.1570$.
Upper and lower figures are obtained with $g_V^2(\pi/a)$ and
$g_V^2(1/a)$, respectively.
The solid lines represent the results of the linear fit.
For $f^+(q^2)$, $\chi^2/$dof are 
$1.9$ and $4.1$ for $q^*=\pi/a$ and $1/a$ respectively.
$\chi^2/$dof for $f^0(q^2)$ are less than $0.5$
in the both cases. } 
\label{fig:FF02renorm}
\end{figure}

\clearpage

\begin{figure}[p]
\begin{center}
\vspace*{-0.8cm}
\leavevmode\psfig{file=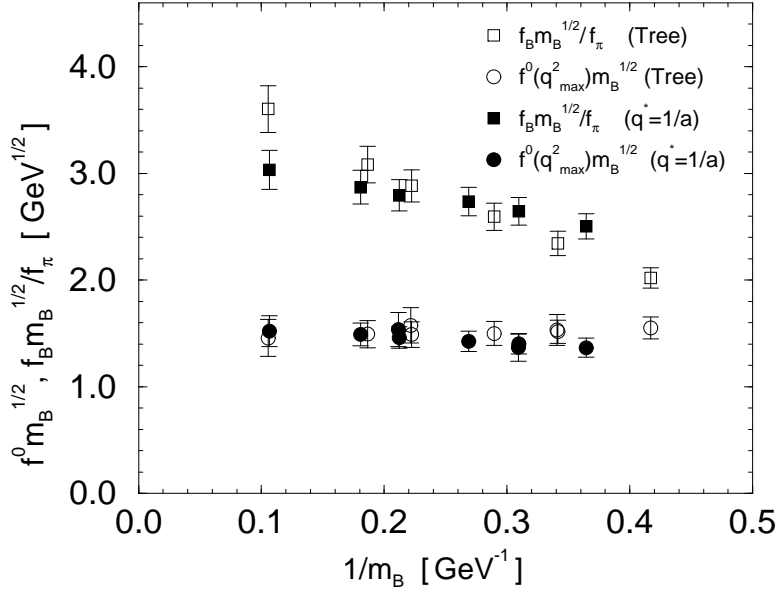,width=\figwidth}
\vspace{-1cm}
\end{center}
\caption{ 
Comparison of $f^0(q^2_{\rsub{max}})$ with $f_B/f_{\pi}$
multiplying the factor $\sqrt{m_B}$ in the chiral limit.
Open and filled symbols are at the tree level and at the one-loop with 
$g_V^2(1/a)$, respectively. }
\label{fig:SPT}
\end{figure}

\begin{figure}[p]
\begin{center}
\vspace*{-0.8cm}
\leavevmode\psfig{file=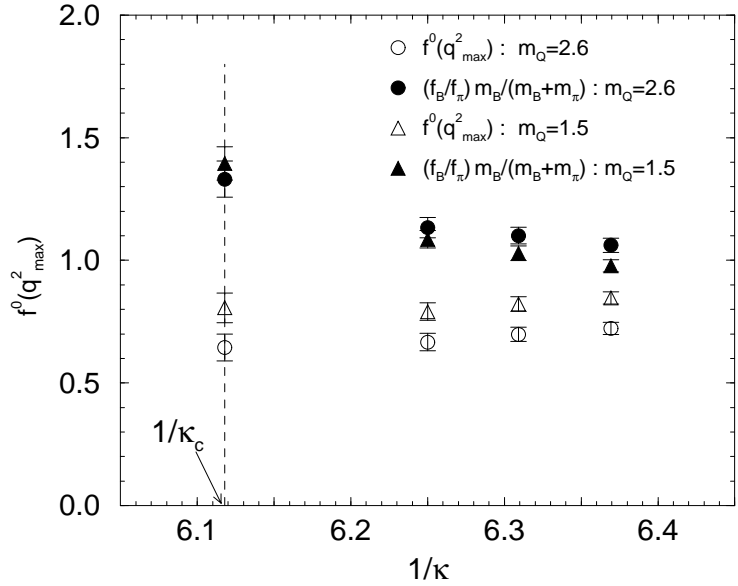,width=\figwidth}
\vspace{-1cm}
\end{center}
\caption{ 
$f^0(q^2_{\rsub{max}})$ and $(f_B/f_{\pi})m_B/(m_B+m_{\pi})$ 
for light quark masses.
Two cases of $m_Q$, $2.6$ and $1.5$, are displayed 
at the tree level. }
\label{fig:SPT_ext}
\end{figure}

				    \end{document}